\documentclass[aps,reprint,twocolumn,superscriptaddress,citeautoscript,showpacs,amsmath,amssymb,floatfix,prb,nobalancelastpage]{revtex4-1}
\usepackage{graphicx,epsfig,units,array}
\usepackage{xcolor} 
\usepackage{amsfonts,mathrsfs,amsbsy,bm}
\usepackage{textcomp,gensymb} 
\usepackage{float}
\usepackage{multirow}
\newcolumntype{C}{>{\centering\arraybackslash}p{1.6cm}}

\begin{document}

\title{Competing magnetic phases and itinerant magnetic frustration in SrCo$_{2}$As$_{2}$}

\author{Bing~Li}
\affiliation{Ames Laboratory, Ames, IA, 50011, USA}
\affiliation{Department of Physics and Astronomy, Iowa State University, Ames, IA, 50011, USA}

\author{B.~G.~Ueland}
\affiliation{Ames Laboratory, Ames, IA, 50011, USA}

\author{W.~T.~Jayasekara}
\affiliation{Ames Laboratory, Ames, IA, 50011, USA}
\affiliation{Department of Physics and Astronomy, Iowa State University, Ames, IA, 50011, USA}

\author{D.~L.~Abernathy}
\affiliation{Oak Ridge National Laboratory, Oak Ridge, TN, 37831, USA}

\author{N.~S.~Sangeetha}
\affiliation{Ames Laboratory, Ames, IA, 50011, USA}

\author{D.~C.~Johnston}
\affiliation{Ames Laboratory, Ames, IA, 50011, USA}
\affiliation{Department of Physics and Astronomy, Iowa State University, Ames, IA, 50011, USA}

\author{Qing-Ping Ding}
\affiliation{Ames Laboratory, Ames, IA, 50011, USA}
\affiliation{Department of Physics and Astronomy, Iowa State University, Ames, IA, 50011, USA}

\author{Y.~Furukawa}
\affiliation{Ames Laboratory, Ames, IA, 50011, USA}
\affiliation{Department of Physics and Astronomy, Iowa State University, Ames, IA, 50011, USA}

\author{P.~P.~Orth}
\affiliation{Ames Laboratory, Ames, IA, 50011, USA}
\affiliation{Department of Physics and Astronomy, Iowa State University, Ames, IA, 50011, USA}

\author{A.~Kreyssig}
\affiliation{Ames Laboratory, Ames, IA, 50011, USA}
\affiliation{Department of Physics and Astronomy, Iowa State University, Ames, IA, 50011, USA}

\author{A.~I.~Goldman}
\affiliation{Ames Laboratory, Ames, IA, 50011, USA}
\affiliation{Department of Physics and Astronomy, Iowa State University, Ames, IA, 50011, USA}

\author{R.~J.~McQueeney}
\affiliation{Ames Laboratory, Ames, IA, 50011, USA}
\affiliation{Department of Physics and Astronomy, Iowa State University, Ames, IA, 50011, USA}

\begin{abstract}
Whereas magnetic frustration is typically associated with local-moment magnets in special geometric arrangements, here we show that SrCo$_{2}$As$_{2}$ is a candidate for frustrated itinerant magnetism. Using inelastic neutron scattering (INS), we find that antiferromagnetic (AF) spin fluctuations develop in the square Co layers of SrCo$_{2}$As$_{2}$ below $T\approx100$~K centered at the stripe-type AF propagation vector of $(\frac{1}{2},~\frac{1}{2})$, and that their development is concomitant with a suppression of the uniform magnetic susceptibility determined via magnetization measurements.  We interpret this switch in spectral weight as signaling a temperature-induced crossover from an instability towards FM ordering to an instability towards stripe-type AF ordering on cooling, and show results from Monte-Carlo simulations for a $J_{1}$-$J_{2}$ Heisenberg model that illustrate how the crossover develops as a function of the frustration ratio $-J_1/(2J_2)$.  By putting our INS data on an absolute scale, we quantitatively compare them and our magnetization data to exact-diagonalization calculations for the $J_{1}$-$J_{2}$ model  [N.~Shannon \textit{et al.}, Eur.~Phys.~J.~B \textbf{38}, 599 (2004)], and show that the calculations predict a lower level of magnetic frustration than indicated by experiment.  We trace this discrepancy to the large energy scale of the fluctuations ($J_{\text{avg}}\agt75$~meV), which, in addition to the steep dispersion, is more characteristic of itinerant magnetism.
\end{abstract}

\date{\today}
\maketitle
\section{Introduction}

Itinerant magnetism originates from the properties of band electrons near the Fermi surface, rather than localized valence electrons associated with an atomic magnetic moment.  A common example is Stoner ferromagnetism (FM), which is driven by the combination of high electronic density-of-states per magnetic atom at the Fermi energy $D(E_{\text{F}})$ and strong electronic-correlation energy $I$.  When the Stoner parameter is large, $ID(E_{\text{F}}) \gg1$, spontaneous itinerant FM order occurs, such as that found in Co, Fe, and Ni at rather high Curie temperatures ($T_{\text{C}}>600$~K) \cite{Kittel96, Nagaoka66, Tasaki93}.  On the other hand, weak itinerant FM, such as ZrZn$_{2}$, have $ID(E_{\text{F}}) \gtrsim 1$, characteristically low values for $T_{\text{C}}$, and smaller saturated moments \cite{Wohlfarth68}. Stoner paramagnets (PM), such as Pd \cite{Mueller70}, with $ID(E_{\text{F}}) \lesssim 1$, are nearly FM and have an enhanced uniform magnetic susceptibility~\cite{Liu_1979}.

Superconductivity exists in the midst of stripe-type antiferromagnetic (AF) fluctuations in various iron-pnictide superconductors \cite{Dai12,Johnston10,Canfield10}; however, many structurally related but nonsupercondcucting cobalt pnictides are considered to be weak itinerant FM.  For example, LaCo$_{2}$P$_{2}$ \cite{Reehuis94} is a metallic FM with a small saturation moment relative to the Curie-Weiss effective moment (i.e.\ a large Rhodes-Wohlfarth parameter \cite{Rhodes63, Santiago17}).  Tetragonal CaCo$_{2}$P$_{2}$ \cite{Reehuis98} and CaCo$_{2-y}$As$_{2}$ \cite{Cheng12, Quirinale13, Jayasekara17} have long-range A-type AF order, with an ordered magnetic moment of $\mu<0.5~\mu_{\text B}/\text{Co}$, consisting of two-dimensional ($2$D) FM square Co layers coupled by much weaker AF interlayer interactions.  Thus, in these two compounds the strong intralayer FM is predominant.

\begin{figure}
	\includegraphics[width=1.0\linewidth]{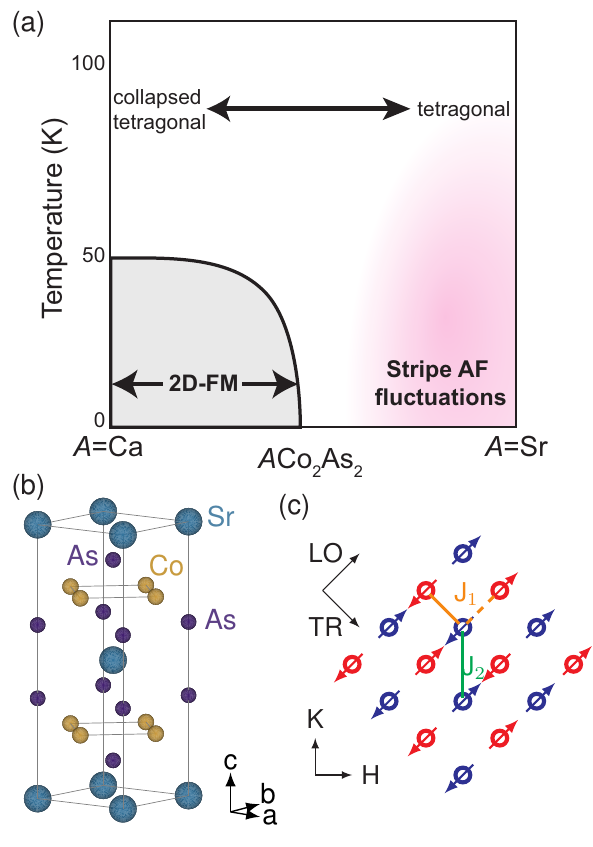}
	\caption{(a) Schematic magnetic phase diagram showing the evolution from stripe-type antiferromagnetic (AF) spin fluctuations to two-dimensional ferromagnetic ($2$D-FM) order in (CaSr)Co$_{2}$As$_{2}$. (b) The body-centered-tetragonal unit cell of SrCo$_{2}$As$_{2}$ with each square-Co sublattice indicated. (c) Diagram showing stripe-type AF order and the nearest-neighbor ($J_{1}$) and next-nearest-neighbor ($J_2$) magnetic interactions in the $\mathbf{HK}$ reciprocal-lattice plane. Red and blue symbols denote the two AF sublattices, and the transverse (TR) and longitudinal (LO) directions referred to in the text are labeled.  $J_{1}$ is FM and $J_{2}$ is AF, and the orange dashed line illustrates a frustrated $J_{1}$ exchange path.  \textsc{vesta} \cite{Momma11} was used to generate (b).}
	\label{fig_J1J2}
\end{figure}

On the other hand, the related compounds BaCo$_{2}$As$_{2}$ \cite{Sefat09}, SrCo$_{2}$P$_{2}$ \cite{Jia09}, and SrCo$_{2}$As$_{2}$ \cite{Pandey13} present more of a mystery.  These materials have large estimated Stoner parameters that should be sufficient for FM ordering, but long-range magnetic order does not occur. An enhanced magnetic susceptibility in these materials may be interpreted as evidence for Stoner PM, and could explain the lack of magnetic order.  However, the discovery via inelastic neutron scattering (INS) of relatively strong low-temperature AF spin fluctuations in SrCo$_{2}$As$_{2}$ centered at reciprocal-lattice momenta $\mathbf{Q}=\mathbf{Q}_{\text{stripe}}$ corresponding to an AF propagation vector for the square-Co planes of $\bm{\tau}_{\text{stripe}}=(\frac{1}{2},\frac{1}{2})$ is very surprising \cite{Jayasekara13}.

An investigation of solid solutions of (Ca,Sr)Co$_{2}$As$_{2}$ \cite{Jayasekara13,Ying13, Sangeetha17} and (Ca,Sr)Co$_{2}$P$_{2}$ \cite{Jia09} demonstrate tunability from 2D-FM to stripe-type AF fluctuations, but long-range stripe-type AF order is never observed in either of these series \cite{Sangeetha17, Jia09}.  On the other hand, recent data for Sr$_{1-x}$La$_{x}$Co$_{2}$As$_{2}$ show that replacing as little as $2.5\%$ Sr by La induces FM order \cite{Shen18}, suggesting that SrCo$_{2}$As$_{2}$  is close to an instability towards a FM phase.  Recent INS experiments have also found FM spin fluctuations in SrCo$_{2}$As$_{2}$, but the reported results do not include a detailed temperature dependence of the fluctuations \cite{Li_2019}.

Figure~\ref{fig_J1J2}(a) shows a schematic magnetic phase diagram for (Ca,Sr)Co$_{2}$As$_{2}$, and Fig.~\ref{fig_J1J2}(b) shows the $I4/mmm$ unit cell of the compounds.  The area in the phase diagram labeled $2$D-FM indicates a region encompassing three AF order phases.  Each AF phase has FM-aligned square Co planes stacked AF, with the periodicity of the stacking and the direction of the ordered magnetic moment distinguishing each phase \cite{BLi_2019}.

The competition between stripe-type AF and FM phases within a single Co-As plane may be captured using a local-moment $J_1$-$J_2$ Heisenberg model for a square magnetic lattice with a spin $\mathbf{S}_{i}$ ($\mathbf{S}_{j}$) at site $i$($j$):
\begin{equation}
\mathcal{H}=J_{1}\sum_{\text{NN}}\mathbf{S}_{i}\cdot\mathbf{S}_{j}+J_{2}\sum_{\text{NNN}}\mathbf{S}_{i}\cdot\mathbf{S}_{j},\label{eq_J1J2_Ham}
\end{equation}
where $J_1$ and $J_2$ are the nearest-neighbor (NN) and next-nearest-neighbor (NNN) exchange, respectively. Figure~\ref{fig_J1J2}(c) shows a Co plane with the $J_1$ and $J_2$ exchange paths labeled, and arrows indicate what stripe-type AF order would look like if it existed in SrCo$_{2}$As$_{2}$.  Since the interlayer coupling is weak compared to $J_1$ and $J_2$ \cite{Jayasekara13, BLi_2019}, we can safely ignore it for our analysis.

The quotient $-J_{1}/(2J_{2})$ can be identified as the frustration ratio, which quantifies the level of magnetic frustration present.  In particular, competing NN FM exchange ($J_{1}<0$) and NNN AF exchange ($J_{2}>0$) may cause either FM [$-J_{1}/(2J_{2})>1$] or stripe-type AF order [$-J_{1}/(2J_{2})<1$] in the $T=0$ ground state. However, extreme geometric frustration [$-J_{1}/(2J_{2}) \approx1$] can suppress long-range order and lead to spin-liquid behavior \cite{Shannon04}.  For example, in the presence of FM $J_1$ and AF $J_2$, the lack of long-range order may be a consequence of the Co spin's inability to simultaneously satisfy its NN and NNN interactions.  This is shown by the dashed orange line in Fig.~\ref{fig_J1J2}(c), which identifies a frustrated $J_{1}$ pathway.   Importantly, for CaCo$_{2-y}$As$_{2}$, which has $-J_{1}/(2J_{2})\approx1$, the frustration ratio manifests directly in the spin-excitation spectrum, where ridges of scattering appear in INS data \cite{Sapkota17}.  The ridges are a signature of the frustrated magnetism, and are observed instead of the magnon spectrum expected for the A-type AF order.

The magnetism  of Fe-pnictide superconductors, and, more generally, of a frustrated square lattice has also been approached using  itinerant magnetic models \cite{Mizusaki06,Yamada13, Han09}.  Interestingly, the calculated magnetic phase diagrams agree with those determined using the $J_{1}$-$J_{2}$ local-moment Heisenberg model, albeit within certain limits.   This dual character of the magnetism has been explored in other Fe-pnictide materials \cite{Xu_2008,Han09, Wyosocki_2011,Yamada13,Glasbrenner_2015}.  In particular, Ref.~[\onlinecite{Han09}] reports results from first-principle density-functional-theory calculations which show that the in-plane magnetic interactions are short ranged and can be effectively described in terms of NN and NNN exchange constants.

In this paper, we reveal through INS data for $\chi(\mathbf{Q}_{\text{stripe}}, E)$, where $E$ is energy, that the stripe-type AF fluctuations found in SrCo$_{2}$As$_{2}$ at $T=5$~K weaken but do not become broader in $\mathbf{Q}$ with increasing temperature.  This suggests that the associated fluctuating magnetic moment becomes suppressed with increasing temperature without a concurrent shrinking of the magnetic correlation length. As the fluctuations diminish, we show that a peak in the dc magnetic susceptibility $\chi(\mathbf{Q}=\bm{0},E=0)\equiv M/H$ develops, where $M$ is the magnetization and $H$ is the applied magnetic field.

Through comparison of our experimental data to results from our own classical Monte-Carlo (MC) simulations and exact-diagonalization calculations from Ref.~[\onlinecite{Shannon04}] using Eq.~\eqref{eq_J1J2_Ham} with $S=1/2$, we show that the switch in spectral weight from $\mathbf{Q}=\bm{0}$ to $\mathbf{Q}_{\text{stripe}}$ upon cooling signals a crossover from the compound being close to an instability towards FM ordering to being close to an instability towards stripe-type AF ordering.  This implies that the stripe-type AF and FM phases lie close in total energy, and we find that the frustration ratio is almost twice as large as that expected from comparing the anisotropy of the AF fluctuations observed via INS to the dc magnetic susceptibility, Monte-Carlo, and exact-diagonalization results.  We interpret the enhanced level of frustration as being due to the large energy scale of the spin-fluctuations, which we associate with the itinerancy of SrCo$_{2}$As$_{2}$'s magnetism.  

\section{Methods}
\subsection{Experiment}
Single crystals of SrCo$_{2}$As$_{2}$ were grown from solution using Sn flux and their compositions were verified as described in Ref.~[\onlinecite{Pandey13}].   Measurements of $M$ were made on a single-crystal sample between $T=1.8$ and $300$~K using a Quantum Design, Inc., Magnetic Properties Measurement System (MPMS). High-temperature magnetization measurements between $T=300$ and $900$~K were performed using the vibrating sample magnetometer (VSM) option of a Quantum Design, Inc., Physical Properties Measurement System (PPMS).  The magnetization measurements determined $\chi(\bm{0}, 0)$.  

INS measurements were made on the Wide Angular-Range Chopper Spectromenter (ARCS) \cite{Abernathy12} at the Spallation Neutron Source at Oak Ridge National Laboratory. Eleven single crystals of SrCo$_2$As$_2$ with a total mass of $3.12$~g were co-aligned with their $(H, H, L)$ planes lying horizontal, where the momentum transfer is given as  $\mathbf{Q}=(2\pi/a)H\hat{\mathbf{i}}+(2\pi/a)K\hat{\mathbf{j}}+(2\pi/c)L\hat{\mathbf{k}}$ and $a=3.95$~\AA\ and $c=11.8$~\AA\ are the lattice parameters.  Rocking scans of the co-aligned assembly gave full-widths at half-maximum of less than $2\degree$.  The $\mathbf c$ axis was kept fixed along the direction of the incident neutron beam,  and incident neutron energies of $E_{\text{i}}=75$ and $250$~meV were used. Data were recorded at $T=50$, $100$, $150$, and $200$~K.  Data at $T= 5 $~K have been reported previously, but in arbitrary units \cite{Jayasekara13}.  INS data shown in this report are normalized by the incoherent scattering of vanadium and corrected for the sample temperature in order to obtain the imaginary part of the dynamical magnetic susceptibility, $\chi^{\prime\prime}(\mathbf{Q}, E)$, in absolute units of $\mu_{\text{B}}^{2}/$eV-fu, where fu stands for formula unit.

Ultra-low temperature nuclear magnetic resonance (NMR) measurements of $^{59}$Co  ($I$ = $\frac{7}{2}$, $\frac{\gamma_{\rm N}}{2\pi} = 10.03$~MHz$/$T)  and $^{75}$As  ($I$ = $\frac{3}{2}$, $\frac{\gamma_{\rm N}}{2\pi} = 7.2919$~MHz$/$T) were conducted down to $T=0.05$~K on a single-crystal sample of SrCo$_2$As$_2$  using a lab-built phase-coherent spin-echo pulse spectrometer with an Oxford dilution refrigerator.  The $^{75}$As-NMR and $^{59}$Co-NMR spectra were obtained by sweeping a magnetic field applied perpendicular to the $\mathbf{c}$ axis at a fixed frequency of $49.5$~MHz.  The temperature dependence of the ac susceptibility $\chi_{\text{ac}}$ was effectively measured down to $T=0.05$~K under $H=0$~T by measuring the NMR coil tank circuit resonance frequency $f$.  $f$ is associated to $\chi_{\text{ac}}$ by $f=1/2\pi\sqrt{L_0(1+\chi_{\text{ac}})C}$, where $L_0$ is the inductance without a sample present. 

\subsection{Simulation}
We performed Monte-Carlo simulations of the classical $J_1$-$J_2$ model on a $L \times L$ square lattice with a linear size of $L=32$ or $64$ over a total of $2.048 \times 10^8$ MC steps. Each MC step consisted of a Metropolis update, a heat-bath update \cite{Miyatake86}, and a parallel-tempering step \cite{Swendsen86}. The systems were simulated at $50$ different temperatures using a geometric spacing between $0.01 < k_{\text{B}}T/|J_1| < 3$ in parallel, where $k_{\text{B}}$ is the Boltzmann constant.  Errors were computed using the Jackknife method over $1024$ equally spaced measurements (every $10^5$ MC steps). Measurements of the simulated systems were taken after an initial thermalization period of $1.024 \times 10^8$ MC steps.

\section{Results}
\subsection{Magnetic susceptibility at $\mathbf{Q}\bm{=0}$}\label{sec_chi_Q0}

\begin{figure}
	\includegraphics[width=1.0\linewidth]{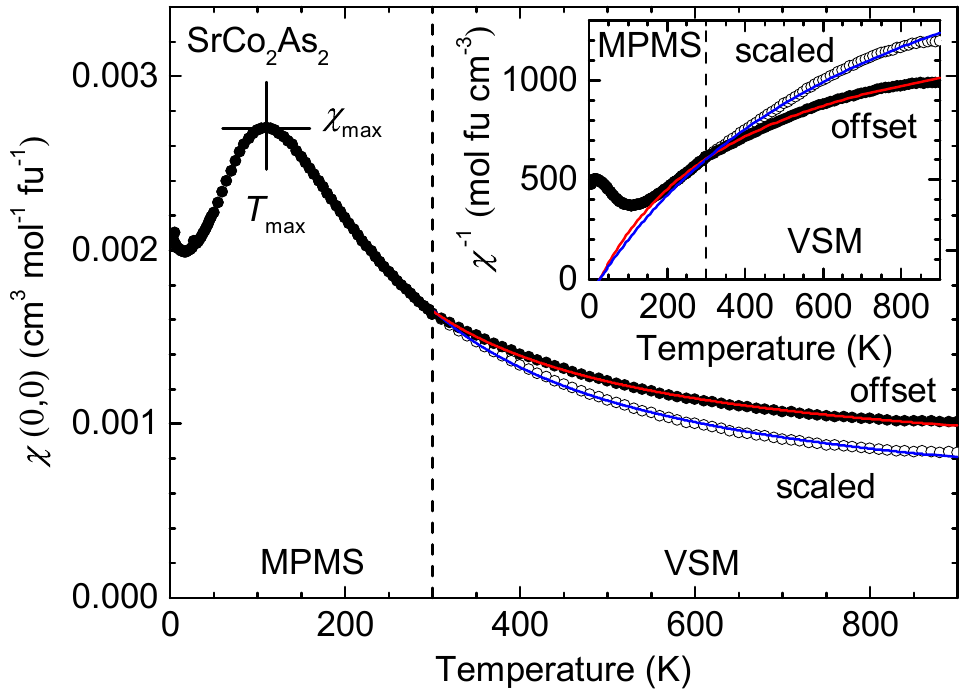}
	\caption{Magnetic susceptibility $\chi(\mathbf{Q}=\bm{0},E=0)\equiv M/H$ of SrCo$_{2}$As$_{2}$ as a function of temperature for a magnetic field of $H=3$~T.  Data for $T\le300$~K are  from MPMS measurements and data for $T\ge300$~K are from VSM measurements.  The maximum in $\chi(T)$ is indicated by \mbox{$\chi_{\text{max}}=0.027$ cm$^{3}/$mol-fu} and $T_{\text{max}}=110(5)$~K.  The inset shows $\chi^{-1}(T)$.  VSM data were both offset (filled circles) and scaled (empty circles) to join the MPMS data at $T=300$~K, as described in the text.  The red and blue lines show fits to Eq.~\eqref{CW_eq}.} 
	\label{CW_fig}
\end{figure}

Figure \ref{CW_fig} displays $\chi(\mathbf Q=\bm{0},E=0)$ for $T=2$ to $900$~K and $H=3$~T, for which a maximum is visible at $T_{\text{max}}=110$(5) K with $\chi_{\text{max}}=0.027$ cm$^{3}/$mol-fu.  Data between $T=300$ and $900$~K allow for determination of the Curie-Weiss (CW) parameters \cite{Kittel96} well above $T_{\text{max}}$ through fits to
\begin{equation}
\chi(\mathbf{Q}=\bm{0},E=0)=\chi_{0}+\frac{C}{T-\theta},
\label{CW_eq}
\end{equation}
where $C$ is the Curie constant, $\theta$ is the Weiss temperature, and $\chi_{0}$ is the temperature-independent susceptibility.

\begin{table}
	\caption {Results from fits of Eq.~\eqref{CW_eq} to the high-temperature VSM data.  Fitting errors for the parameters are given.  For the average of the scaled and offset fit parameters, the error is obtained from the difference between the two.}
	\renewcommand\arraystretch{1.25}
	\begin{ruledtabular}
		\begin{tabular}{ c | c | c | c }
			
			~  ~             								&  ~scaled ~ 		&  ~offset ~       &~average~     \\
			\hline
			$\chi_{0}$ (10$^{-4}$~cm$^{3}/$mol-fu) 	& $4.22(14)$ 		& $6.89(11)$ 	& $5.5(1.3)$ \\
			$C$~(cm$^{3}/$K-mol-fu) 							&  $0.337(14)$ & $0.263(11)$ & $0.30(4)$ \\
			$\theta$~(K) 													& $27(9)$ 		& $27(9)$ 		& $27(9)$ \\
			$\mu_{\text{eff}}=\sqrt{8C/2}$~($\mu_{\text{B}}/$Co) & $1.16(3)$ & $1.03(2)$ & $1.10(6)$   \\
			
		\end{tabular}
	\end{ruledtabular}
	\label{CW_table}
\end{table}

The MPMS ($T\le300$~K) and VSM ($T\ge300$~K) data do not join smoothly due to calibration issues with the VSM thermometry, so we compared two methods for joining the data: (1) adding a constant offset and (2) multiplying by a scale factor.  The VSM data were fit by Eq.~\eqref{CW_eq} for each method, with the results being given in Table \ref{CW_table}.  Figure~\ref{CW_fig} also shows that $\chi(\bm{0},0)$ levels off to a large value at high $T$, which gives a value for $\chi_{0}$  consistent with the Pauli susceptibility estimated from the density of states at the Fermi level of $D(E_{\text{F}})=11$~states$/$eV-fu \cite{Pandey13}:
\begin{equation}
\begin{split}
\chi_{0} \approx \chi_{\text{Pauli}}&=\mu_{\text{B}}^{2}D(E_{\text{F}})\\*& = 3.5 \times 10^{-4}~\text{cm}^{3}/\text{mol-fu}.
\end{split}
\end{equation}
Fits performed to our MPMS data over $T=200$ to $300$~K yielded parameters similar to those reported in Ref.~[\onlinecite{Pandey13}].

\subsection{Magnetic susceptibility at $\mathbf{Q}\bm{=}\mathbf{Q}_{\text{stripe}}$}\label{chi_Qstripe}

The imaginary part of the magnetic susceptibility is calculated from the INS data according to
\begin{equation}
\begin{split}
\chi^{\prime\prime}&(\mathbf{Q}, E)=\\*&\frac{2\pi}{(\gamma r_{\text{0}})^{2}}\frac{S(\mathbf{Q}, E)-S_{\text{bkgd}}(\mathbf{Q}, E)}{f^{2}(\text{Q})}({1-e^{-E/k_{\text{B}}T}}), \label{eq_chi_INS}
\end{split}
\end{equation}
where $S(\mathbf{Q}, E)$ is the scattering intensity, $S_{\text{bkgd}}(\mathbf{Q}, E)$ is an isotropic nonmagnetic background, $(\gamma r_{\text{0}})^{2}=290.6$~mbarn$/$sr, and $f(\text{Q})$ is the magnetic form factor of the Co$^{2+}$ ion. The nonmagnetic background was estimated by a procedure similar to the one used in Ref.~[\onlinecite{Tucker12}]. To summarize, the magnetic scattering intensity near $\mathbf{Q}_{\text{stripe}}=(\pm0.5, \pm0.5)$ was masked.  Then, data points with the same values of $\sqrt{H^{2}+K^{2}}$ (within a tolerance of $0.025$~rlu) and energy transfer $E$ (within a tolerance of the step size in $E$ after reduction of the time-of-flight data) were averaged to form $S_{\text{bkgd}}(\mathbf{Q}, E)$.

\begin{figure}
	\includegraphics[width=1.0\linewidth]{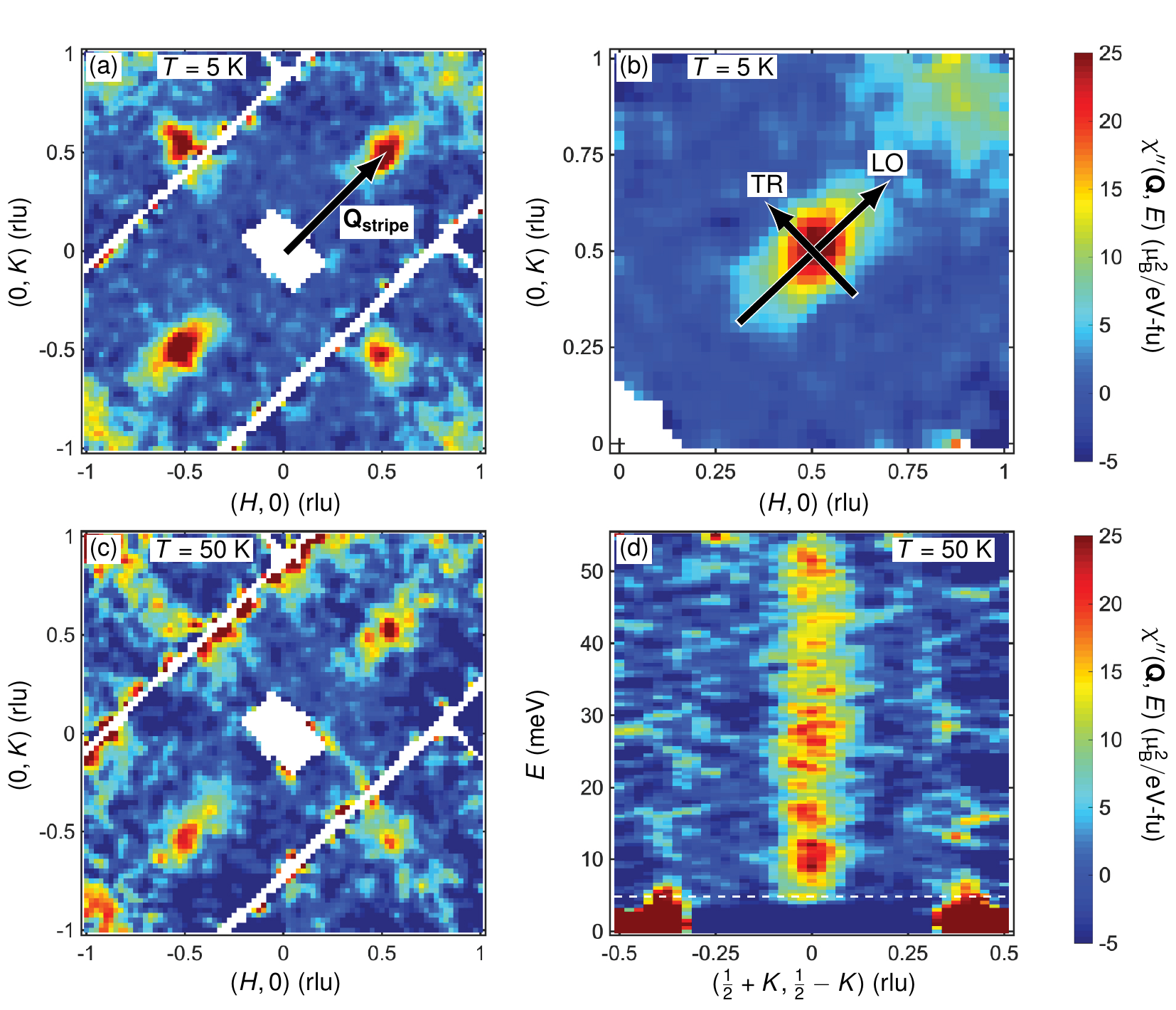}
	\caption{The imaginary part of the magnetic susceptibility $\chi^{\prime\prime}$ in absolute units from inelastic neutron scattering data showing the presence of anisotropic stripe-type AF spin fluctuations in SrCo$_{2}$As$_{2}$ at $T=5$~K [(a),(b)] and $50$~K[(c),(d)].  A background has been subtracted as described in the text.  [(a),(c)] Scattering in the $(H,~K)$-plane averaged over a neutron energy transfer range of $E=5$ to $20$ meV. (b) Data as in (a), but averaged over symmetry-equivalent quadrants.  Anisotropy is clearly visible, with the scattering being more extended along the longitudinal (LO) direction than along the transverse (TR) direction.  (d) The steeply dispersing behavior of the spin fluctuations as seen for the TR direction.  These data are averaged over $\pm0.1$~rlu in the LO direction.  Data below $E=5$~meV (dashed white line) are contaminated by strong elastic scattering.  All data are for an incident neutron energy of $E_{\text{i}}=75$~meV.}
	\label{INST5T50}
\end{figure}
\subsubsection{Weakening of the stripe-type spin fluctuations with increasing temperature}
Figure \ref{INST5T50} gives an overview of the INS due to anisotropic spin fluctuations centered at $\mathbf{Q}_{\text{stripe}}=(1/2,1/2,L)$ for $T=5$~K [Figs.~\ref{INST5T50}(a) and \ref{INST5T50}(b)] and $50$~K [Figs.~\ref{INST5T50}(c) and \ref{INST5T50}(d)].  Since the INS measurements were made with the sample's $\mathbf{c}$ axis parallel to the incoming beam, the measured value of $L$ depends on $E$.  Thus, summing over a range of $E$ corresponds to summing over a range in $L$.  Previous data show that the spin fluctuations centered at $\mathbf{Q}_{\text{stripe}}$ only weakly disperse along $\mathbf{L}$ \cite{Jayasekara13}, making them quasi-$2D$ and predominately governed by the intralayer NN and NNN exchange.

Figures~\ref{INST5T50}(a) and ~\ref{INST5T50}(b) demonstrate the reciprocal-space anisotropy of the spin fluctuations:  they are broad in the longitudinal (LO) direction ($\parallel\mathbf{Q}_{\text{stripe}}$) and narrow in the transverse (TR) direction ($\perp\mathbf{Q}_{\text{stripe}}$).  Figure~\ref{INST5T50}(c) shows that the fluctuations are still present at $T=50$~K but are weaker than at $5$~K. The temperature dependence of the anisotropy is quantified by making cuts across the INS scattering peaks along the LO and TR directions, examples of which are given in Fig.~\ref{fig_tdep}. (See also Fig.~\ref{cuts} in Appendix~\ref{INS_diff_anal}).  The peak widths  $\kappa_{\text{LO}}$ and $\kappa_{\text{TR}}$ in the cuts determine the anisotropy parameter $\eta$:
\begin{equation}
\eta= \frac{\kappa_{\text{TR}}^{2}-\kappa_{\text{LO}}^{2}}{\kappa_{\text{TR}}^{2}+\kappa_{\text{LO}}^{2}}. \label{eq_eta}
\end{equation}
$\eta$  is $-0.5$ at $T=5$~K and $-0.6$ at $50$~K. Within the random-phase approximation (RPA) to the $J_{1}$-$J_{2}$ model, it can be shown [see Appendix \ref{MF_scale}, equation \eqref{etaJ1J2}] that
\begin{equation}
-\eta=-\frac{J_{1}}{2J_{2}}. \label{eq_eta_J1J2}
\end{equation}
Thus $\eta$ serves as a measure of the frustration ratio \cite{Sapkota17}.  

\begin{figure}[]
	\includegraphics[scale=0.9]{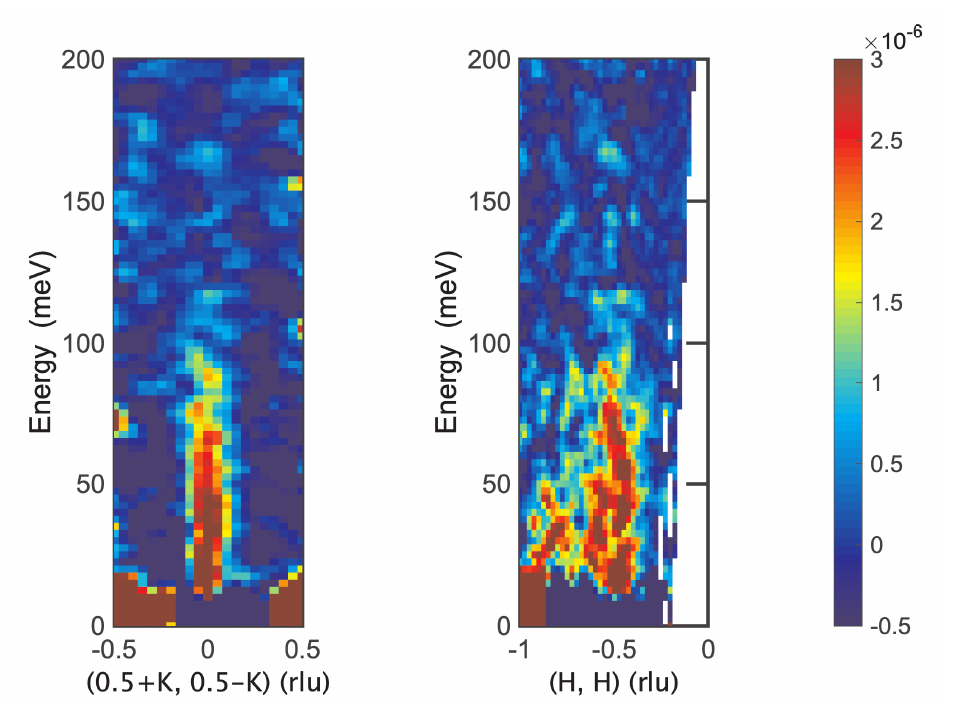}
	\caption{(a) Transverse (TR) and (b) longitudinal (LO) slices of background-subtracted inelastic neutron scattering data for an incident neutron energy of $E_{\text{i}}=250$~meV. The TR slice is averaged over $\pm$0.1 rlu in the LO direction, the LO slice is averaged over $\pm$0.1 rlu in the TR direction.  The intensity is given in arbitrary units.}
	\label{250meV}
\end{figure}

Figures~\ref{INST5T50}(d), \ref{250meV}(a), and \ref{fig_tdep}(a) show the steep dispersion of the spin fluctuations in the TR direction, whereas Fig.~\ref{250meV}(b) shows the weaker dispersion in the LO direction.  Figure~\ref{250meV} further shows that the fluctuations extend up to $E=100$~meV, with no clear sign of broadening in $\mathbf Q$ with increasing $E$.  Rather, the dispersion is more reminiscent of that seen for itinerant magnets \cite{Chatterji06}.

Given the steep dispersion, we can only obtain a lower bound for the magnitude of the transverse velocity 
\begin{equation}
\begin{split}
v_{\text{TR}} &= \frac{\Delta E}{\Delta q} \\*&\gtrsim \frac{50\ \text{meV}}{0.2\ \text{\AA}^{-1}} = 250\ \text{meV}\ \text{\AA},
\end{split}\label{eq_vtr}
\end{equation}
where $\Delta q$ is the distance away from $\mathbf{Q}_{\text{stripe}}$.  As shown in Appendix~\ref{INS_Jav}, this leads to a lower bound for the average value of the exchange energy of
\begin{equation}
\begin{split}
J_{\text{avg}}&=\sqrt{J_{1}^{2}+J_{2}^{2}}\\&\approx75~\text{meV}.
\end{split}\label{eq_Javg}
\end{equation}
 
\begin{figure}[H]
	\includegraphics[width=1\linewidth]{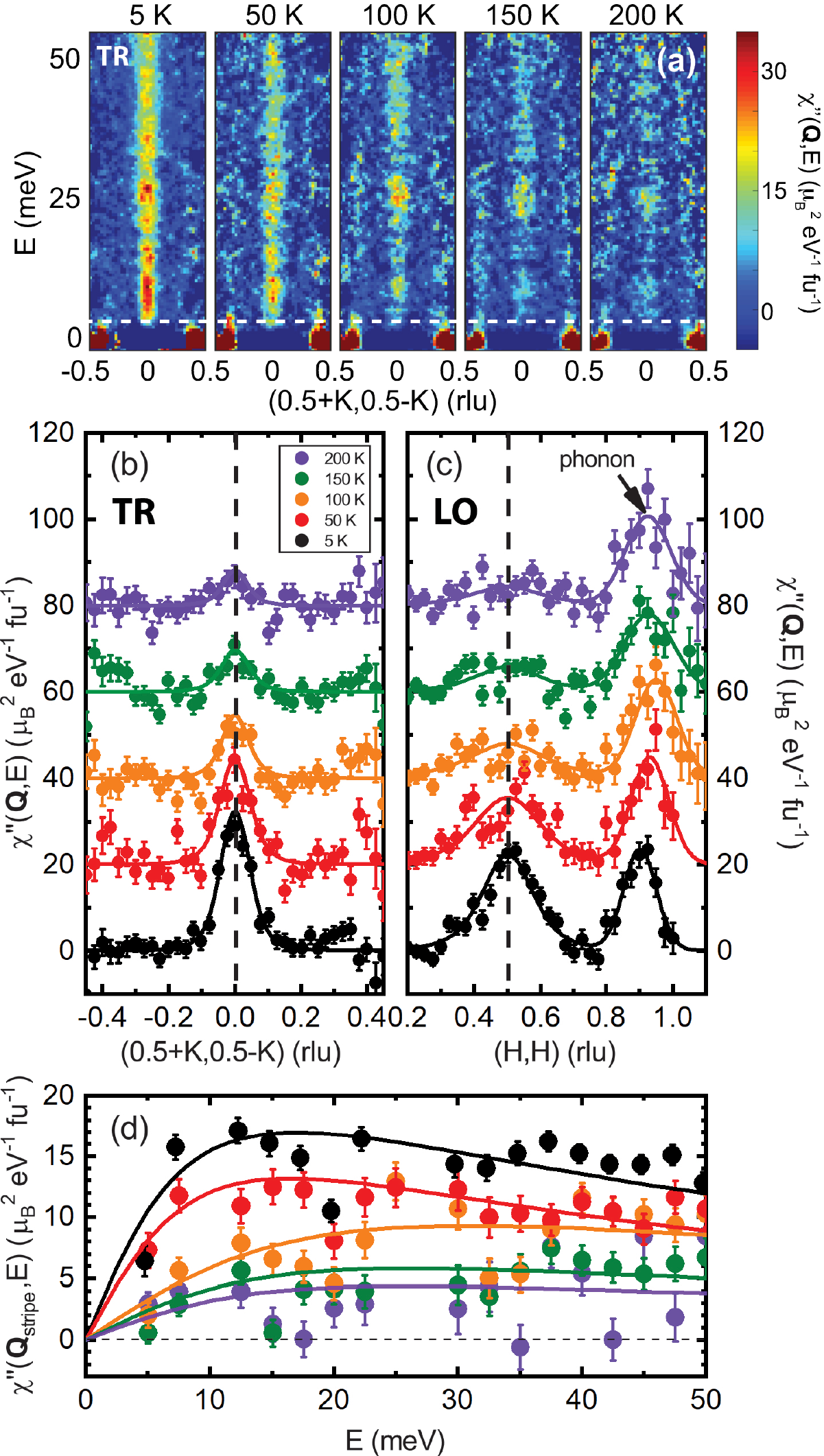}
	\caption{ (a) Color image of the dynamical susceptibility obtained from inelastic neutron scattering data  through $(\frac{1}{2},~\frac{1}{2},~L)$ and $E$ in absolute units of $\mu_{\text{B}}^{2}/$eV-fu for the TR direction. Slices are averaged over $\pm 0.1$ rlu in LO direction. $L$ is  tied to $E$ due to the sample's $\mathbf{c}$ axis being oriented along the incoming beam.  Data below $E\approx 5$\ meV (dashed white line) are contaminated by large elastic scattering. (b) TR and (c) LO cuts of the dynamical susceptibility through $\mathbf{Q}_{\text{stripe}}$ (dashed line) averaged over $\pm$0.1 rlu and $E=10$ to $15$~meV at different temperatures as listed. (d) Energy dependence of the spin fluctuations at $\mathbf{Q}_{\text{stripe}}$ for different temperatures.  The color scheme in (d) is the same as in (b) and (c).  All data are for an incident neutron energy of $E_{\text{i}}=75$~meV.} 
	\label{fig_tdep}
\end{figure}

Figure \ref{fig_tdep}(a) shows the suppression of the spin fluctuations with increasing temperature in more detail, and Figs.~\ref{fig_tdep}(b) and \ref{fig_tdep}(c) show TR and LO cuts averaged over $E=10$ to $15$~meV for each temperature measured.  The peak in Fig.~\ref{fig_tdep}(c) located near $(0.9, 0.9)$~rlu is due to phonon contamination.  Figure~\ref{fig_tdep}(d) demonstrates the suppression of $\chi^{\prime\prime}(\mathbf{Q}_{\text{stripe}}\pm\mathbf{q},E)$ versus $E$ with increasing temperature. 

A key observation is that the stripe-type AF spin fluctuations weaken with increasing temperature, whereas the peak widths are not strongly affected.  This suggests a suppression of the fluctuating AF moment rather than the reduction of the spin-spin correlation length generally expected for a local-moment magnet as $T$ is increased further away from the magnetic-ordering temperature. 

To understand these temperature-dependent changes, we fit $\chi^{\prime\prime}(\mathbf{Q}_{\text{stripe}},E)$ at each temperature to a diffusive model for the spin fluctuations based on the local-moment $J_{1}$-$J_{2}$ Heisenberg Hamiltonian given in Eq.~\eqref{eq_J1J2_Ham}.  We discuss this model below.

\subsubsection{Fits to a diffusive model within a random-phase approximation to the $J_1$-$J_2$ model}
The diffusive model \cite{Diallo10, Inosov09, Sapkota17} within a RPA to the $J_{1}$-$J_{2}$ model yields an imaginary susceptibility:
\begin{equation}
\begin{split}
\chi^{\prime\prime}(\mathbf{Q}_{\text{stripe}}+\mathbf{q},E)&=\\*&\frac{\chi^{\prime}(\mathbf{Q}_{\text{stripe}},0)\Gamma_{T}E}{\Gamma_{T}^{2}[1+\xi_{T}^{2}(q^{2}+2\eta q_{x}q_{y})]^{2}+E^2},
\end{split}
\label{eq_diffusive}
\end{equation}
where $\chi^{\prime}(\mathbf{Q}_{\text{stripe}},0)$ is the staggered susceptibility at $\mathbf{Q}_{\text{stripe}}$, $\Gamma_{T}$ is the relaxation rate, $\xi_{T}$ is the correlation length, and $\eta=J_{1}/(2J_{2})$ is the reciprocal-space anisotropy of the spin fluctuations.  The subscripts $x$ and $y$ correspond to  perpendicular directions connecting NN Co.

\begin{figure}
	\includegraphics[width=1.0\linewidth]{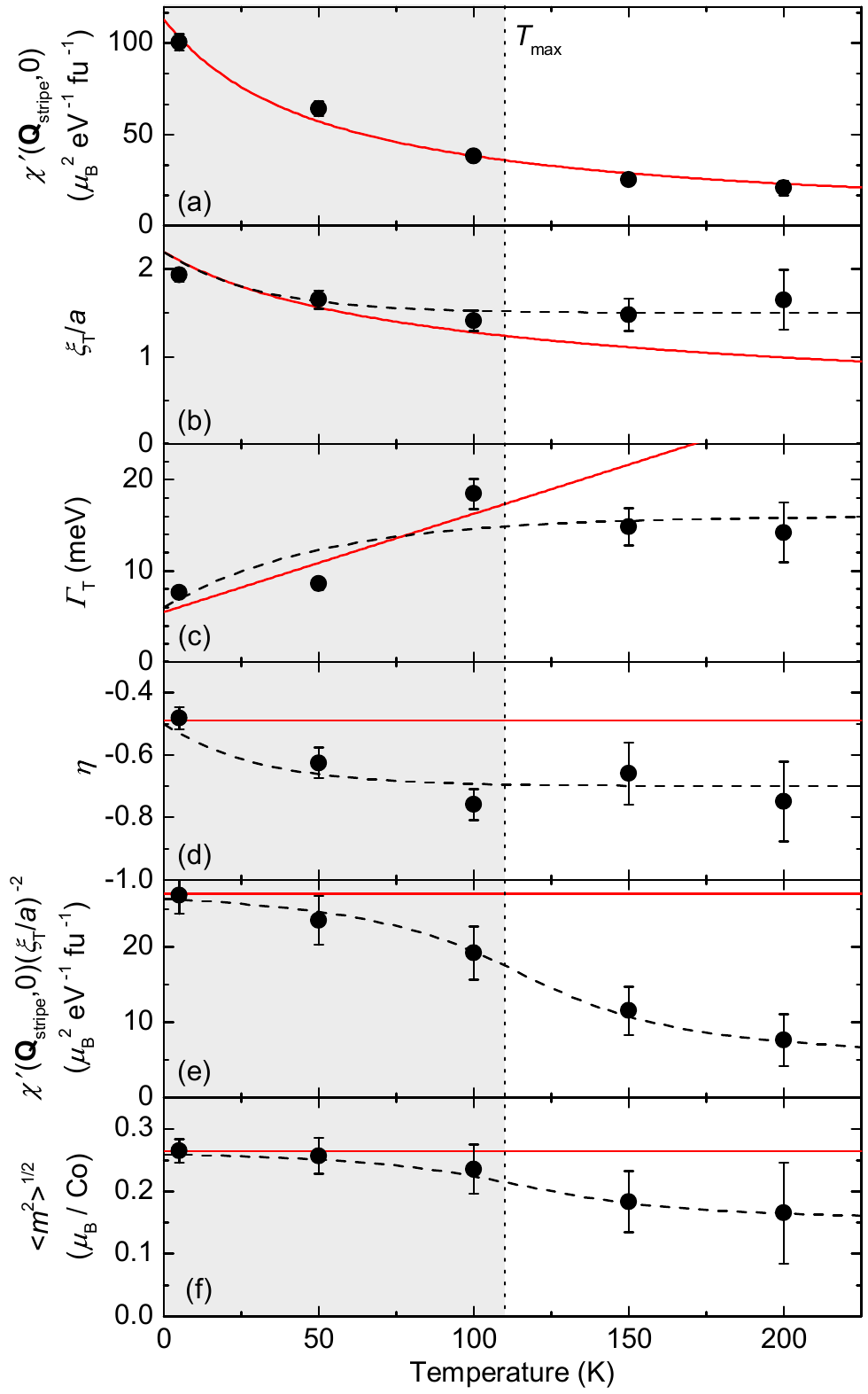}
	\caption{Temperature dependencies of parameters determined from fits of inelastic neutron scattering data to the diffusive model imaginary susceptibility given in Eq.~\eqref{eq_diffusive}. Red lines indicate expectations from critical behavior for the $J_{1}$-$J_{2}$ model in a random-phase approximation (see Appendix~\ref{MF_scale}), and black dashed lines are guides to the eye. (a) The staggered susceptibility at $\mathbf{Q}_{\text{stripe}}$.  The red line shows a Curie-Weiss fit. (b)--(f) The AF correlation length $\xi_{T}/a$ (b),  relaxation rate $\Gamma_{T}$ (c), reciprocal space anisotropy $\eta$ (d), scaling relation $\chi^{\prime}(\mathbf{Q}_{\text{stripe}},0)(\xi_{T}/a)^{-2}$ (e), and fluctuating AF moment $\langle m^{2} \rangle$ (f).   Parameters in the shaded area (below $T_{\text{max}}$) show reasonable scaling for incipient stripe-type AF order.}
	\label{fig_fits}
\end{figure}

TR and LO cuts through $\mathbf{Q}_{\text{stripe}}$ for energy transfer ranges of $E = 5$ to $10$, $10$ to $15$, $30$ to $40$, and $40$ to $50$~meV, where the magnetic scattering largely avoids phonon scattering, are shown in Fig. 12 in Appendix~\ref{INS_diff_anal}.  Together with the energy dependence of $\chi^{\prime\prime}(\mathbf{Q}_{\text{stripe}},E)$ shown in Fig.~\ref{fig_tdep}(d), the cuts were simultaneously fit by Eq.~\eqref{eq_diffusive}.  The temperature dependence of the fitted parameters are shown in Fig.~\ref{fig_fits}.

Appendix~\ref{MF_scale} shows that $\chi^{\prime}(\mathbf{Q}_{\text{stripe}},E=0)$  in Fig.~\ref{fig_fits}(a) may be fit to the form: 
\begin{equation}
\chi^{\prime}(\mathbf{Q}_{\text{stripe}},E=0) = \chi_{s}\frac{|T_{\text{N}}|}{T-T_{\text{N}}}, \label{eq_E0_RPA_Curie}
\end{equation}
 where $T_{\text{N}}$ is the N\'{e}el temperature.  This gives a bare staggered susceptibility of $\chi_{s}=110(20)\ \mu_{\text{B}}^{2}/$eV-fu, an effective staggered moment of $\mu_{\text{eff}}=\sqrt{3k_{\text{B}}|T_{\text{N}}|\chi_{s}}=0.85\ \mu_{\text{B}}/$Co, and $T_{\text{N}}=-51(7)$~K. Since long-range AF order does not occur, $T_{\text{N}}$ is negative.

Figure \ref{fig_fits}(b) shows that $\xi_{T}$ is weakly dependent on temperature and does not conform to the expected scaling behavior for our RPA-based diffusive model (see Appendix~\ref{MF_scale}) of
\begin{equation}
\xi_{T}^{2} \sim \xi_{0}^{2}\frac{|T_{\text{N}}|}{T-T_{\text{N}}},
\end{equation}
as for $T>100$~K the correlation length remains constant. Figure \ref{fig_fits}(c) also shows that the expected critical behavior for the relaxation rate:
\begin{equation}
\Gamma_{T} \sim \frac{\gamma(T-T_{\text{N}})}{|T_{\text{N}}|},
\end{equation}
where $\gamma$ is the Landau damping, arising from the itinerancy of the material, fits poorly above $T=100$~K.  The overall breakdown of critical behavior is best illustrated by Fig.~\ref{fig_fits}(e), which demonstrates that the scaling quantity $\chi^{\prime}(\mathbf{Q}_{\text{stripe}},0)(\xi_{T}/a)^{-2}$ varies with temperature.  Equation~\eqref{C14} shows that this quantity should be constant in $T$ for our diffusive model.

The fluctuating AF moment,
 \begin{equation}
\left<m^2\right>=\frac{1}{2}\frac{3}{\pi}\frac{\int\chi^{\prime\prime}(\mathbf{Q},E)(1-e^{-E/k_\text{B}T})^{-1}\text d\mathbf{Q}\text dE}{\int\text d\mathbf{Q}},
\label{moment}
\end{equation}
was determined by integration of Eq.~\eqref{eq_diffusive} up to a cutoff energy of $E=100$~meV after substituting the fitted parameters.   The factor of $\frac{1}{2}$ in Eq.~\eqref{moment} takes into consideration that there are two Co atoms per fu, and the range of integration over $\mathbf{Q}$ is $(0\le Q_x\le\frac{2\pi}{a},~0\le Q_y\le\frac{2\pi}{a})$.

The temperature dependence of $\left<m^{2}\right>$  is plotted in Fig.~\ref{fig_fits}(f), which shows that it decreases above $T\approx100$~K.  Overall, Fig.~\ref{fig_fits} demonstrates that the stripe-type AF spin fluctuations in SrCo$_{2}$As$_{2}$ follow the critical behavior expected for the diffusive model reasonably well for $T\alt100$~K, even though the compound never attains long-range stripe-type AF order. 

\subsection{Nuclear magnetic resonance}

\begin{figure}[]
	\centering
	\includegraphics[width=1.0\linewidth]{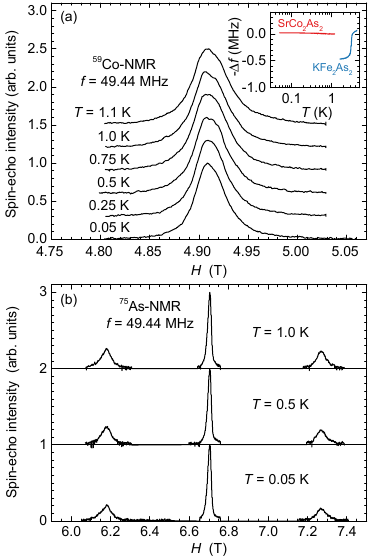}
	\caption{(a) $^{59}$Co-NMR spin-echo data for SrCo$_{2}$As$_{2}$ versus magnetic field for various temperatures.  (inset)  Temperature dependence of the change in the NMR coil tank circuit resonance frequency $-\Delta f$ for SrCo$_2$As$_2$ (red curve).  The blue curve is from Ref.~[\onlinecite{Wiecki18}] and shows  $-\Delta f(T)$ for the superconductor KFe$_2$As$_2$, for which there is a clear anomaly at $T_{\text{c}}=3.3$~K.  (b)  $^{75}$As-NMR spin-echo data for SrCo$_{2}$As$_{2}$ versus magnetic field for various temperatures.  The spin-echo intensities are given in arbitrary units.}
	\label{NMR_combo}
\end{figure}

Previously reported data for SrCo$_{2}$As$_{2}$ have demonstrated that no magnetic or superconducting phase transitions occur down to $T=1.8$~K \cite{Pandey13, Wiecki15}.  To examine if a phase transition occurring below $T=1.8$~K is related to the decrease in $\chi(T)$ below $T_{\text{max}}$, we made ac susceptibility and NMR measurements down to $0.05$~K.

The inset to Fig.~\ref{NMR_combo}(a) shows the temperature dependence of the shift in resonance frequency of the NMR tank circuit $-\Delta f(T)$ for either SrCo$_2$As$_2$ or the superconductor KFe$_2$As$_2$ placed within the pickup coils.  It demonstrates that $-\Delta f(T)$ for KFe$_2$As$_2$ shows a sharp change at its superconducting transition temperature of $T_{\text{ c}} = 3.3$~K \cite {Wiecki18}, which is due to diamagnetic shielding, whereas the data for SrCo$_2$As$_2$ show no such feature for $T$ down to $0.05$~K.  Upon taking into consideration previous results for $T\ge1.8$~K \cite{Pandey13, Wiecki15}, these data exclude a superconducting transition occurring for SrCo$_{2}$As$_{2}$ at $T\ge0.05$~K. 

Figures \ref{NMR_combo}(a) and \ref{NMR_combo}(b) show $^{59}$Co- and $^{75}$As-NMR spin-echo data, respectively, for SrCo$_{2}$As$_{2}$ at various temperatures.  No changes with temperature to the shapes of the spectra are seen, which indicates that no magnetic phase transitions are detected down to $T=0.05$~K.  Further, Fig.~\ref{NMR_combo}(b) shows no abrupt temperature-dependent changes to the spacing between quadruploar-split $^{75}$As-NMR lines.  This likely excludes a structural phase transition as well.

\subsection{Classical Monte-Carlo simulations}

\begin{figure}
	\includegraphics[width=1.0\linewidth]{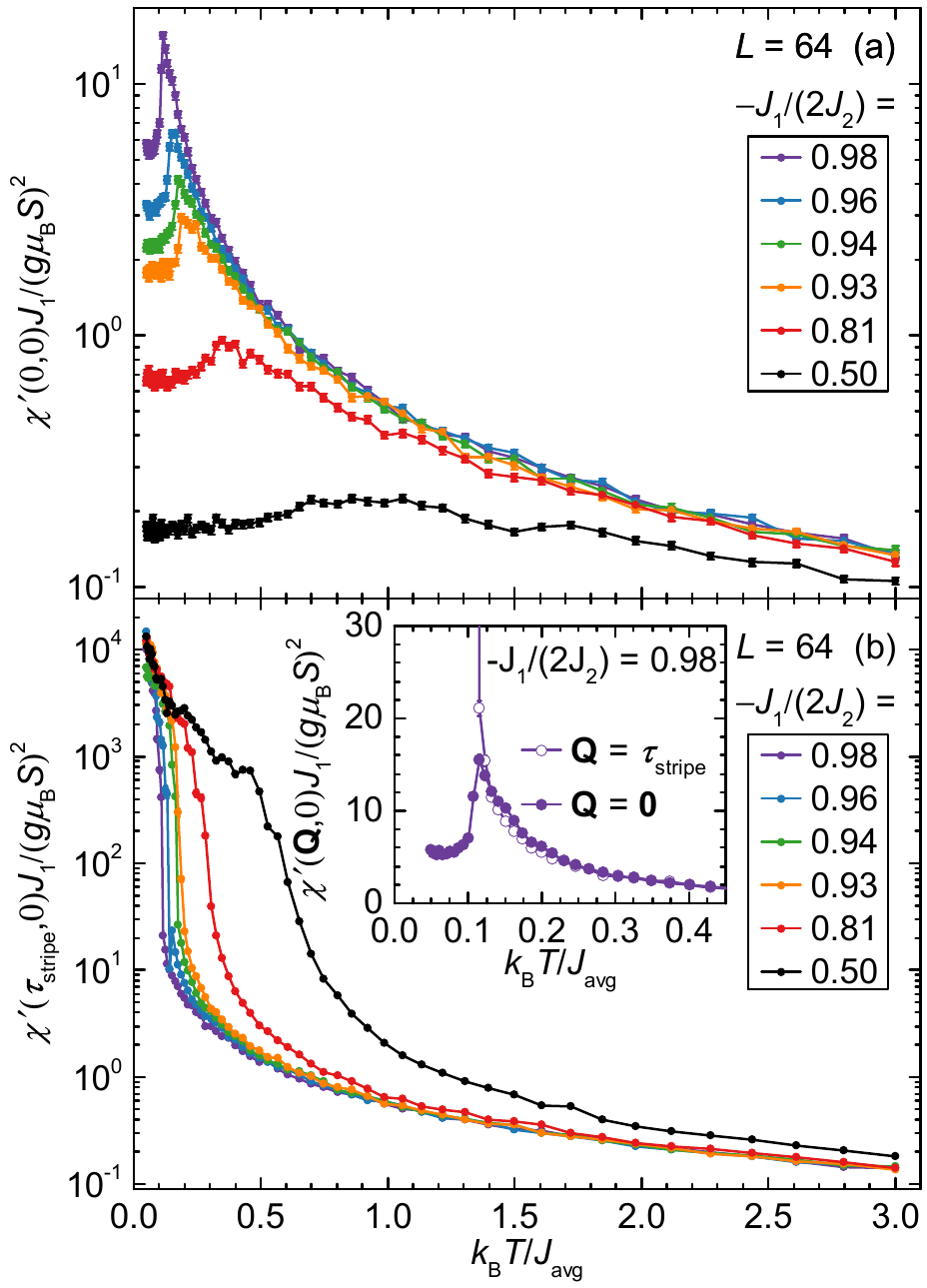}
	\caption{Classical Monte-Carlo simulation results for the $J_1$-$J_2$ model on a $64\times64$ square lattice showing (a) the uniform susceptibility $\chi^{\prime}(\bm{0},0)$ and (b) the staggered susceptibility $\chi^{\prime}(\bm{\tau}_{\text{stripe}} ,0)$.  Data are for various values of $-J_{1}/(2J_{2})$ and are plotted versus an effective temperature $k_{\text{B}}T/J_{\text{avg}}$, where $J_{\text{avg}}=\sqrt{J_{1}^2+J_{2}^{2}}$. Note the logarithmic scales. The inset shows similar classical Monte-Carlo simulation results for $-J_{1}/(2J_{2})=0.98$ for both $\mathbf Q=\mathbf 0$ and $\mathbf Q = \bm{\tau}_{\text{stripe}}$ in linear scale.} 
	\label{fig_mc_sm}
\end{figure}

\begin{figure}
	\includegraphics[width=1.0\linewidth]{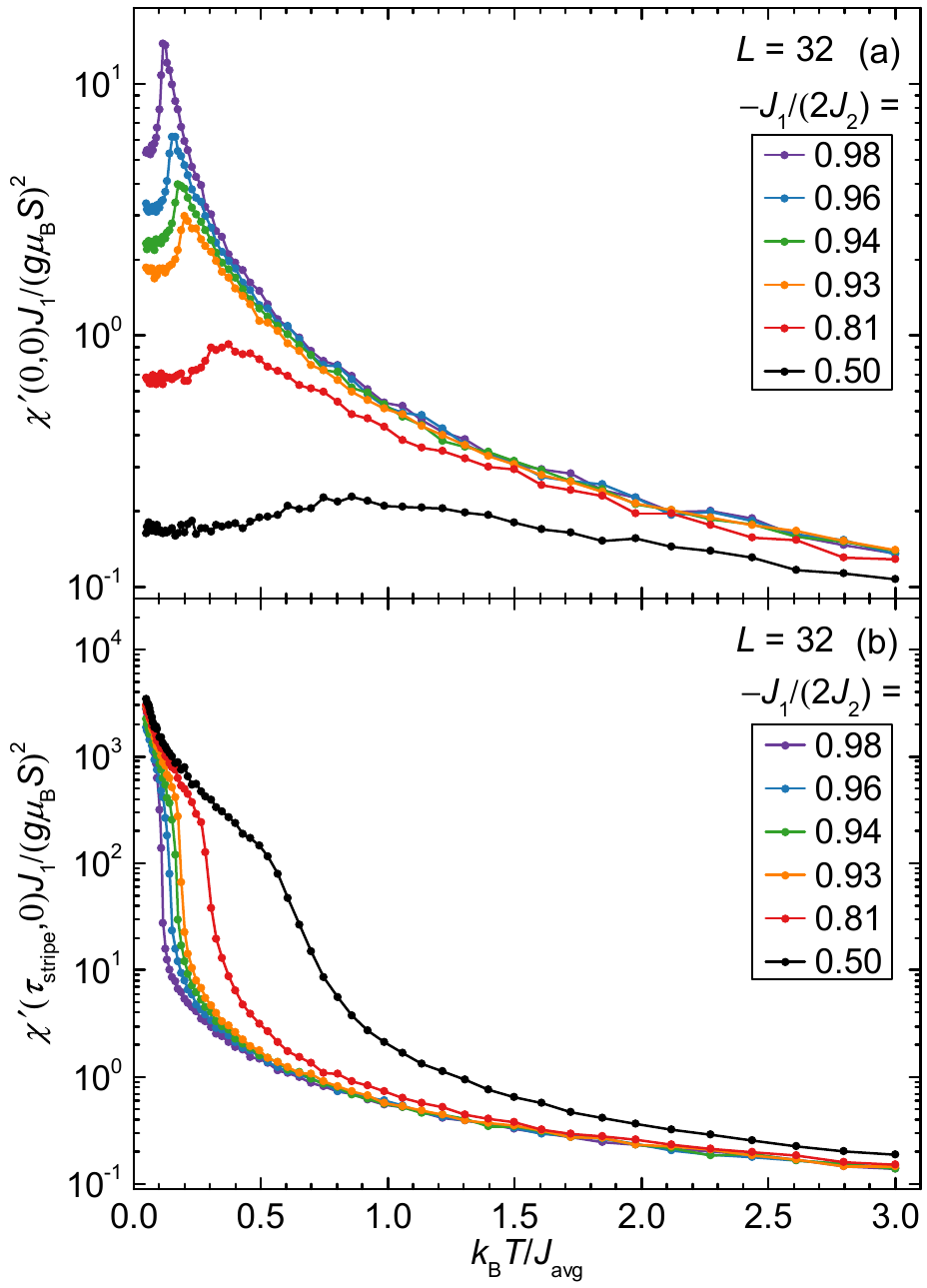}
	\caption{Classical Monte-Carlo simulation results for the $J_1$-$J_2$ model on a $32\times32$ square lattice showing (a) the uniform susceptibility $\chi^{\prime}(\bm{0},0)$ and (b) the staggered susceptibility $\chi^{\prime}(\bm{\tau}_{\text{stripe}} ,0)$.  Data are for various values of $-J_{1}/(2J_{2})$ and are plotted versus an effective temperature $k_{\text{B}}T/J_{\text{avg}}$, where $J_{\text{avg}}=\sqrt{J_{1}^2+J_{2}^{2}}$. Note the logarithmic scales.}
	\label{fig_mc_sm_Jp98}
\end{figure}

To rationalize and interpret the competition between stripe-type AF and FM in the Co-As planes, we have performed large-scale parallel-tempering Monte-Carlo simulations of the $J_1$-$J_2$ model in the classical limit. We set $J_1 < 0$ to be FM and $J_2 > 0$ to be AF, and vary their ratio $0.5 < -J_1/(2J_2) <0.98$. Thus, the ratio goes from the stripe-type AF side of the phase diagram [$-J_{1}/(2J_{2})<1$] towards extreme geometric frustration [$-J_{1}/(2J_{2}) \approx1$].

Figure~\ref{fig_mc_sm}(a) presents the uniform susceptibility $\chi^{\prime}(\bm{0}, 0)$ and Fig.~\ref{fig_mc_sm}(b) gives the staggered susceptibility $\chi^{\prime}(\bm{\tau}_{\text{stripe}} ,0)$  versus $k_{\text{B}}T/J_{\text{avg}}$ calculated for a square lattice with a linear size of $L=64$.  A maximum is evident in $\chi^{\prime}(\bm{0},0)$ which shifts to lower $k_{\text{B}}T/J_{\text{avg}}$ as $-J_{1}/(2J_{2}) \rightarrow1$. This is a signature of the frustration.  On the other hand, $\chi^{\prime}(\bm{\tau}_{\text{stripe}} ,0)$ shows a sharp increase for values of $k_{\text{B}}T/J_{\text{avg}}$ below the value for which $\chi^{\prime}(\bm{0},0)$ has a maximum, and $\chi^{\prime}(\bm{\tau}_{\text{stripe}} ,0)$ grows exponentially below this point due to the divergence of the correlation length as $T\rightarrow0$.

Figure~\ref{fig_mc_sm_Jp98} presents similar data for a square lattice with $L=32$. The positions of the maxima and the values of $\chi^{\prime}(\bm{0} ,0)$ in Figs.~\ref{fig_mc_sm}(a) and \ref{fig_mc_sm_Jp98}(a) show little dependence on $L$.  On the other hand, the values of $\chi^{\prime}(\bm{\tau}_{\text{stripe}} ,0)$ show an obvious $L$ dependence as $k_{\text{B}}T/J_{\text{avg}}\rightarrow0$ in Figs.~\ref{fig_mc_sm}(b) and \ref{fig_mc_sm_Jp98}(b).  This clear dependence of $\chi^{\prime}(\bm{\tau}_{\text{stripe}} ,0)$ on the system-size signals a true divergence in the thermodynamic limit as $T\rightarrow0$, whereas $\chi^{\prime}(\bm{0} ,0)$ is size independent, which implies that the FM fluctuations are not critical.  Rather, they are only enhanced at finite temperature due to the proximity of the nearby FM phase at $-J_1/(2J_2) \agt 1$.
  
 \begin{figure}
 	\includegraphics[width=1.0\linewidth]{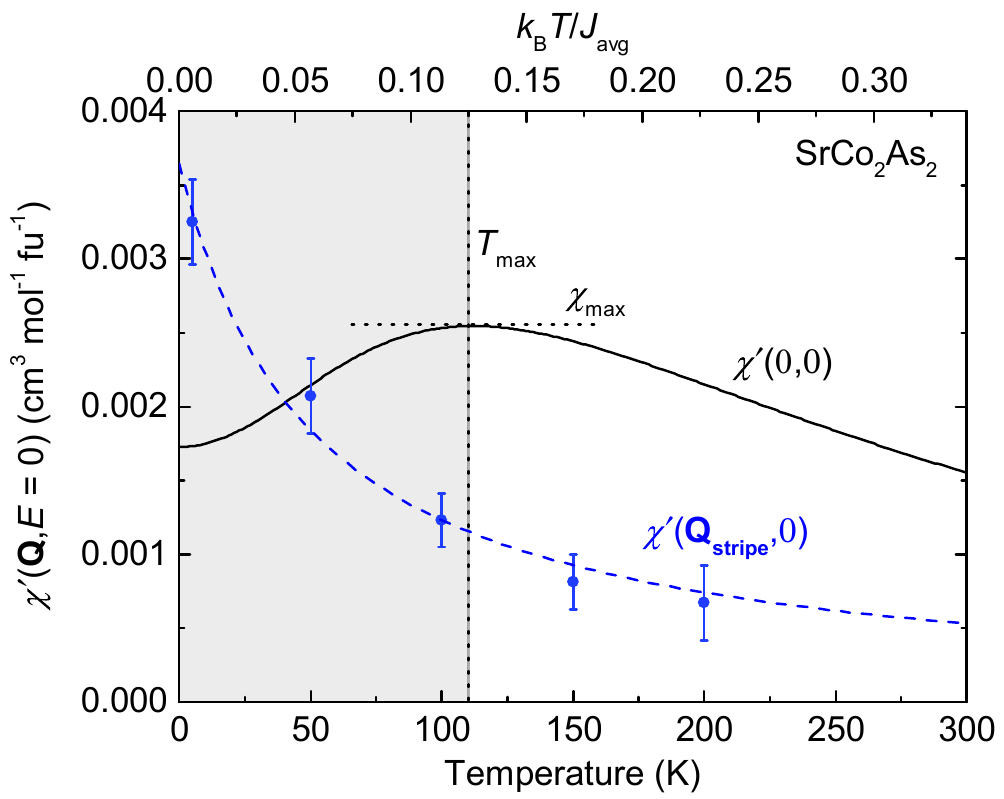}
 	\caption{Fits to the uniform susceptibility $\chi^{\prime}(\mathbf{Q}=\bm{0},E=0)$ of SrCo$_{2}$As$_{2}$ (black line) obtained from Ref.~[\onlinecite{Pandey13}] and the staggered spin susceptibility $\chi^{\prime}(\mathbf{Q}_{\text{stripe}},0)$ obtained from inelastic neutron scattering data (blue symbols).  The dashed blue line is a guide to the eye. Top scale is the effective temperature in units of $k_{\text{B}}T/J_{\text{avg}}.$}
 	\label{chi_dc_INS_comp}
 \end{figure}
  
From our data in Fig.~\ref{CW_fig} and our estimated lower bound for $J_{\text{avg}}$ of $75$~meV from the INS data, we estimate that $k_{\text{B}}T_{\text{max}}/J_{\text{avg}}<0.13$.  This value is approximately reproduced by our MC simulations for $-J_{1}/(2J_{2})=0.98$, data from which are shown in the inset to Fig.~\ref{fig_mc_sm}(b) for $L=64$.   Good qualitative agreement with the experimental data plotted in Fig.~\ref{chi_dc_INS_comp} is seen:  $\chi^{\prime}(\bm{\tau}_{\text{stripe}} ,0)$ steeply increases below the value of $k_{\text{B}}T/J_{\text{avg}}$ for which $\chi^{\prime}(\bm{0},0)$ reaches a maximum at  $k_{\text{B}}T/J_{\text{avg}} \approx 0.125 $.  Nevertheless, the value of $-J_{1}/(2J_{2})$ determined from the INS data is $-\eta=0.5 $ to $0.6$, which is much lower than the value of $0.98$ for the corresponding MC simulations.   Thus, SrCo$_{2}$As$_{2}$ appears to be more frustrated than expected from the measured reciprocal-space anisotropy of the spin fluctuations.

\section{Discussion}

\begin{figure}
	\includegraphics[width=1\linewidth]{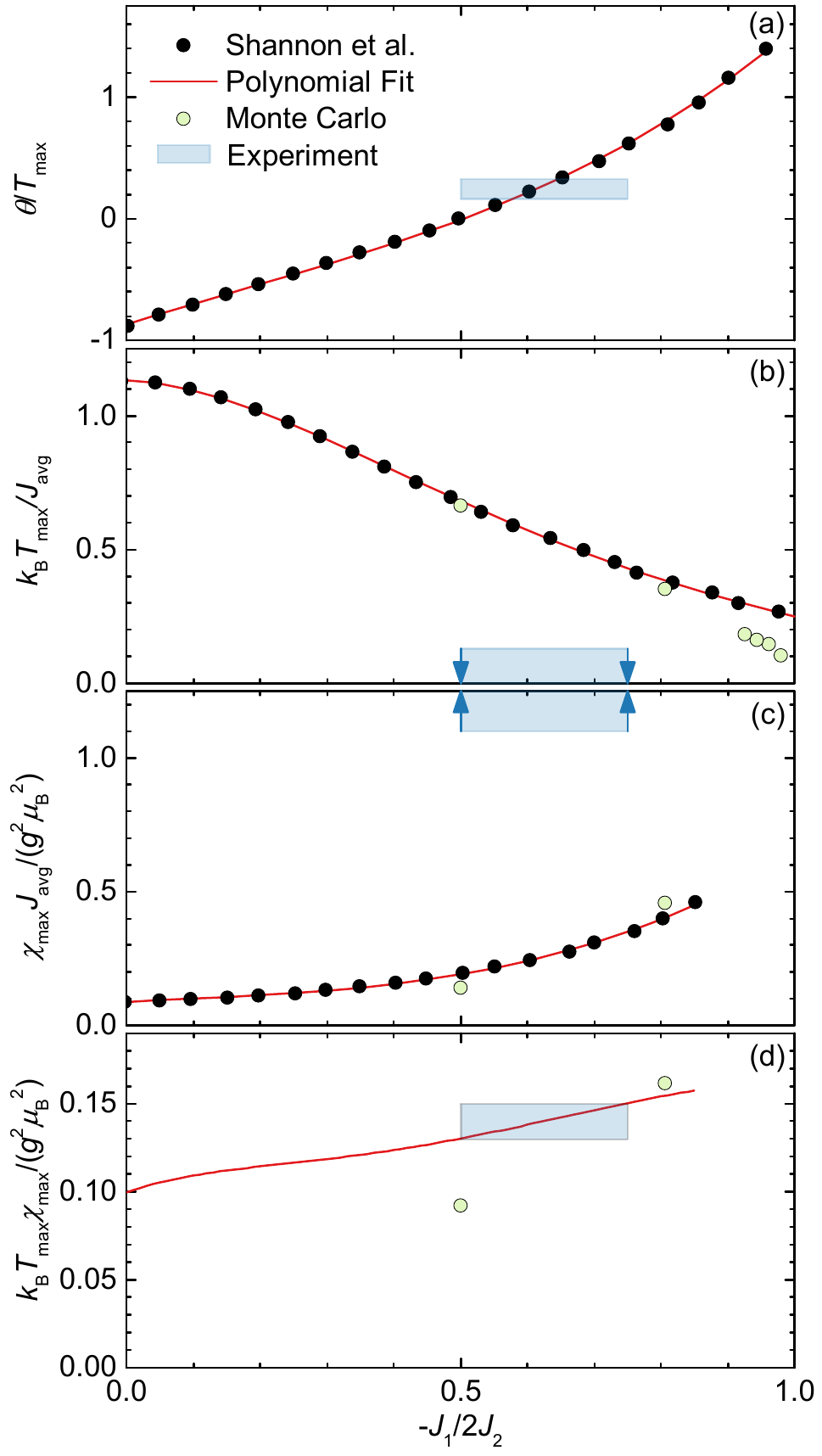}
	\caption{Plots of (a) $\theta/T_{\text{max}}$, (b) $k_{\text{B}}T_{\text{max}}/J_{\text{avg}}$, (c) $\chi_{\text{max}}J_{\text{avg}}/(g^{2}\mu_{\text{B}}^{2})$ and (d)  $k_{\text{B}}T_{\text{max}}\chi_{\text{max}}/(g^2\mu_\text B^2)$ versus the frustration ratio $-J_{1}/(2J_{2})$. Black circles are results from exact-diagonalization calculations given by Shannon et al. in Ref.~[\onlinecite{Shannon04}], and red lines are polynomial fits to the exact-diagonalization results. Black circles with green fill are results from our classical Monte-Carlo simulations. Blue rectangles are parameter ranges determined from experiments.}
	\label{fig_shannon}
\end{figure}

We begin this section by making quantitative comparisons of the measured $\chi(\bm{0},0)$ and INS data to results from exact-diagonalization calculations using the $J_{1}$-$J_{2}$ model described by Eq.~\eqref{eq_J1J2_Ham} with $S=1/2$.  In particular, Shannon et al.  \cite{Shannon04}  report the variations of $\theta/T_{\text{max}}$, $k_{\text{B}}T_{\text{max}}/J_{\text{avg}}$, and $\chi_{\text{max}}J_{\text{avg}}/(g^{2}\mu_{\text{B}}^{2})$ as functions of $-J_{1}/(2J_{2})$.  We have digitized these data and plotted them in Fig.~\ref{fig_shannon}(a), \ref{fig_shannon}(b), and \ref{fig_shannon}(c), respectively. The red curves are polynomial fits to the digitized data. The red curve in Fig.~\ref{fig_shannon}(d) is the product of the fitted red curves in  Figs.~\ref{fig_shannon}(b) and \ref{fig_shannon}(c). Results from our Monte-Carlo simulations are also included as black circles with green fill in Figs.~\ref{fig_shannon}(b), \ref{fig_shannon}(c) and \ref{fig_shannon}(d). Notice that all red curves are dimensionless quantities which can be calculated by theory. 

 Blue rectangles in Fig.~\ref{fig_shannon} are parameter ranges determined from INS and/or magnetization measurements. Their horizontal ranges show that $-J_{1}/(2J_{2})$ = 0.5 to 0.75, as determined from the spatial anisotropy in INS data. As we can only estimate the lower bound of $J_{\text{avg}}$, the blue rectangles in Figs.~\ref{fig_shannon}(b), \ref{fig_shannon}(c) and \ref{fig_shannon}(d) only give bounds for the corresponding parameters. $\theta$, $T_{\text{max}}$, and $\chi_{\text{max}}$ are determined from the magnetization measurement. $g=1.7$ is given in Ref.~[\onlinecite{Pandey13}]. It can be seen that quantities involving values derived from only the magnetization measurement are in good agreement with the exact-diagonalization results, while those involving the value of $J_{\text{avg}}$, determined by INS, are not. This discrepancy can be associated to the large value of $J_{\text{avg}}$.

\begin{table}
	\caption {Comparison of experimental results from inelastic neutron scattering (INS) and dc magnetic susceptibility [$\chi(\bm{0},0)$] measurements with predictions from exact-diagonalization calculations reported in Ref.~[\onlinecite{Shannon04}]  for the $J_{1}$-$J_{2}$ model.  Results from the calculations are shown for frustration ratios of $-J_{1}/(2J_{2})=0.5$ and $0.75$.  $\eta$ and the lower bound for $J_{\text{avg}}$ are obtained from INS data and $k_{\text{B}}T_{\text{max}}$, $\chi_{\text{max}}$, and $\theta$ are obtained from $\chi(\bm{0},0)(T)$ data.   $g=1.7$ is used, which comes from analysis of $\chi(\bm{0},0)$ data  given in Ref.~[\onlinecite{Pandey13}], which uses a value for the spin of $S=1/2$.}
	\renewcommand\arraystretch{1.25}
	\begin{ruledtabular}
		\begin{tabular}{ c | C | C | c }
			\multirow{2}{*}{ } &\multicolumn{2}{c|}{Theory~for~$-J_{1}/(2J_{2})=$} &\multirow{2}{*}{Experiment}     \\
			&$0.5$ & $0.75$  \\
			\hline
			$\eta$ & --- & --- & $-0.63(12)$ \\
			$J_{\text{avg}}$ (meV) & --- & --- & $\agt75$\\
			$T_{\text{max}}$ (K) & --- & --- & $110(5)$ \\
			$\chi_{\text{max}} (\mu_{\text{B}}^{2}/$meV-Co) & --- & --- & $0.043(1)$  \\
			$\theta$ (K) & --- & --- & $27(9)$\\
			\hline
			$\theta/T_{\text{max}}$   & $0$ & $0.61$ &  $0.25(9)$   \\ 
			$k_{\text{B}}T_{\text{max}}\chi_{\text{max}}/(g^{2}\mu_{\text{B}}^{2}  )$ & $0.13$ & $0.15$ & $0.14(1)$\\
			$k_{\text{B}}T_{\text{max}}/J_{\text{avg}}$ & $0.69$ & $0.43$ & $\alt0.13$  \\
			$\chi_{\text{max}}J_{\text{avg}}/(g^{2}\mu_{\text{B}}^{2})$  & $0.19$ & $0.34$ & $\agt1.1$ \\
			
		\end{tabular}
	\end{ruledtabular}
	\label{table_exp_calc}
\end{table}

Table~\ref{table_exp_calc} summarizes the quantities determined from experimental data and exact-diagonalization results. Within these exact-diagonalization results, key indicators of a high degree of frustration are a small value for $k_{\text{B}}T_{\text{max}}/J_{\text{avg}}$ and a large value for $\chi_{\text{max}}J_{\text{avg}}/(g^{2}\mu_{\text{B}}^{2})$. 
In this sense, our experimental measures of the key frustration indicators appear ``more frustrated'' than the $J_{1}$-$J_{2}$ model predicts.  In particular,  $k_{\text{B}}T_{\text{max}}/J_{\text{avg}}$ is much smaller and $\chi_{\text{max}}J_{\text{avg}}/(g^{2}\mu_{\text{B}}^{2})$ is much larger than the values expected from the local-moment model.  This conclusion is supported by the MC results given in Figs.~\ref{fig_mc_sm} and \ref{fig_mc_sm_Jp98}, which show that a value of $-J_1/(2J_2)$ closer to $1$ is more consistent with the measured temperature dependence of the magnetic susceptibility at $\mathbf{Q}=\bm{0}$ and $\mathbf{Q}_{\text{stripe}}$.  SrCo$_{2}$As$_{2}$ is therefore more frustrated than predicted by the value of $\eta=-J_{1}/(2J_{2})$ determined by INS, and the maximum in $\chi^{\prime}(\bm{0},0)$ occurs at a much lower temperature than expected.   This discrepancy is traced to the steep dispersion of the spin fluctuations, and the associated large magnetic energy scale of $J_{\text{avg}}\agt75$~meV, which is more characteristic of an itinerant magnet.

We interpret the suppression of $\chi(\bm{0},0)$ and rise in $\chi(\mathbf{Q}_{\text{stripe}},0)$ below $T_{\text{max}}$ as signaling a crossover from predominantly FM to predominately stripe-type AF fluctuations.  These fluctuations are presumably associated with corresponding FM and AF phases that lie close in energy.  This is supported by the following facts.  First, the magnitude of $\chi(\bm{0},0)$ at high-temperature, the positive Weiss temperature, and the large Stoner parameter of $ID(E_{\text{F}})=2.2$ found in Ref.~[\onlinecite{Pandey13}] are all consistent with a Stoner FM instability.  Second, NMR and INS results both show evidence for FM fluctuations being present \cite{Wiecki18, Li_2019}.  Third, Fig.~\ref{chi_dc_INS_comp} clearly shows that the leading magnetic instability, determined by the maximum in $\chi^{\prime}(\mathbf{Q},0)$, crosses over from $\mathbf{Q}=0$ to $\mathbf{Q}_{\text{stripe}}$ with decreasing temperature.

This scenario of competing FM and stripe-type AF phases is consistent with band structure calculations that find maxima in the generalized electronic susceptibility at both $\mathbf{Q}=\bm{0}$ and $\mathbf{Q}_{\text{stripe}}$ \cite{Jayasekara13}.  Remarkably, even though fluctuations associated with each phase are present at finite temperature and a crossover in the magnetic susceptibility occurs between $\mathbf{Q}=\bm{0}$ and $\mathbf{Q}_{\text{stripe}}$, it is apparently more energetically favorable for the compound to remain paramagnetic. 

Interestingly, Figs.~\ref{fig_mc_sm} and \ref{fig_mc_sm_Jp98} indicate that close to $-J_1/(2 J_2) \alt 1$ FM fluctuations seem to be dominant for a large range of finite $T$ even though the $T=0$ ground state corresponds to stripe-type AF. Previous theory work has shown a similar behavior for AF $J_1$ ($J_1 > 0$) close to $J_1/(2 J_2) = 1$ both in the classical spin limit at $T>0$ \cite{Weber03} as well as in the quantum limit at $T=0$ \cite{Mila91}. These works noted that thermal and quantum fluctuations both favor N\'{e}el-type AF fluctuations for $J_1/(2J_2) \alt 1$ even though the classical ground state at $T=0$ is stripe-type AF.  This leads to a crossover from a high-temperature N\'{e}el-type phase to a low-temperature stripe-type phase  This crossover is similar to our observation for FM $J_1$ of dominant FM fluctuations at large $T$ and a crossing to prevalent stripe-type AF fluctuations at low $T$.

Remarkably, a suppression of $\chi(\bm{0}, 0)$ such as that seen for SrCo$_{2}$As$_{2}$ at $T$ below $110$ K \cite{Pandey13} is a phenomenon seen in some frustrated local-moment square-lattice systems compounds, such as BaCdVO(PO$_{4}$)$_{2}$ \cite{Nath08}.  The unusual behavior of SrCo$_{2}$As$_{2}$ also closely parallels that of a broad class of weak itinerant FMs displaying unusual responses to magnetic fields and temperature which can be characterized as being both itinerant \textit{and} frustrated.  For example, YCo$_{2}$ consists of a geometrically-frustrated corner-shared tetrahedral network of Co ions.  Similar to SrCo$_{2}$As$_{2}$, its high-temperature behavior is consistent with Stoner PM, and upon cooling its $\mathbf{Q}=\bm{0}$ susceptibility reaches a maximum.  Below the temperature of the maximum, the low-energy spin fluctuations become suppressed \cite{Yoshimura88}. Also, similar to the case of (Ca,Sr)Co$_{2-y}$As$_{2}$, whereas YCo$_{2}$ is PM, weak itinerant FM order can be induced in Y(Co$_{1-x}$Al$_{x}$)$_{2}$ for $x>0.11$ \cite{Yoshimura85}.

The application of a magnetic field in the PM state of Y(Co$_{1-x}$Al$_{x}$)$_{2}$ for $x<0.11$ triggers a first-order metamagnetic transition to a FM state that cannot be explained by the alignment of local magnetic moments \cite{Sakakibara86, Sakakibara90}.  This itinerant-electron metamagnetism is proposed to arise from the competition between nearly degenerate PM ground states, one of which is close to a Stoner instability \cite{Yamada93,Takahashi95, Takahashi98}.  Similar observations of high-field metamagnetism, unconventional temperature-dependent uniform magnetic susceptibility, and the evolution of these phenomena upon approach to 2D-FM order in (Ca,Sr)Co$_{2}$P$_{2}$ \cite{Imai14} suggest a close connection between itinerant-electron metamagnetism and itinerant magnetic frustration.

\section{Conclusion}
In summary, we have made temperature dependent INS and magnetization measurements on SrCo$_2$As$_2$ that have determined $\chi(\mathbf{Q},E)$ between $T=5$ and $200$~K. By fitting INS data for $\chi^{\prime\prime}(\mathbf{Q},E)$ to a diffusive model for the $J_1$-$J_2$ Heisenberg Hamiltonian on the square lattice [Eq.~\eqref{eq_diffusive}], we have compared the temperature dependence of $\chi(\mathbf{Q},E=0)$ at $\mathbf{Q}=\bm{0}$, determined via magnetization, to that at $\mathbf{Q}_{\text{stripe}}$.   A decrease in $\chi(\bm{0},0)$ occurs below $T_{\text{max}}=110(5)$~K that is accompanied by a rise in $\chi(\mathbf{Q}_{\text{stripe}},0)$, which signals a shift in magnetic spectral weight from $\mathbf{Q}=\bm{0}$ to $\mathbf{Q}_{\text{stripe}}$.  This occurs despite our NMR data showing that neither FM nor AF order is realized down to $T=0.05$~K.  We interpret the shift as being due to competition between closely lying in-plane FM and stripe-type AF states, which manifests in the observation of steep and anisotropic spin fluctuations centered at $\mathbf{Q}$ corresponding to $\bm{\tau}_{\text{stripe}}$.   Further, within the diffusive model, the anisotropy of the spin fluctuations at $\mathbf{Q}_{\text{stripe}}$ gives a measurement of the level of magnetic frustration: $\eta=J_{1}/(2J_{2})$.

To further understand our data, we have performed classical Monte-Carlo simulations for the $J_1$-$J_2$  model and found that they capture the suppression of $\chi(\bm{0},0)$ and rise in $\chi(\bm{\tau}_{\text{stripe}},0)$ with decreasing temperature. However, the simulation results show that a frustration parameter of $-J_{1}/(2J_{2})\approx0.98$, which is much larger than the range of $-\eta=0.5$ to $0.75$ found by INS, is needed to explain the experimentally determined value for $k_{\text B}T_{\text{max}}/J_{\text{avg}}$.  Upon comparison with previous exact-diagonalization calculations for the $J_{1}$-$J_{2}$ model with $S=1/2$ \cite{Shannon04}, we find that inconsistencies between the experimental data and theory arise due to the large energy scale of the spin fluctuations ($J_{\text{avg}}\agt75$~meV), which, in addition to the steep dispersion observed via INS, is more characteristic of itinerant magnetism.

Thus, we argue that SrCo$_{2}$As$_{2}$ is therefore more frustrated than predicted by the local-moment $J_1$-$J_2$ model due to itinerancy. Remarkably, previous theory results point to similar competition between N\'{e}el- and stripe-type AF states for $-J_{1}/(2J_{2})=-1$ \cite{Mila91,Weber03}.  In addition, the anomalous temperature and magnetic-field responses of other itinerant-electron metamagnetic compounds such as Y(Co$_{1-x}$Al$_{x}$)$_{2}$ \cite{Yoshimura88,Yoshimura85,Sakakibara86, Sakakibara90,Yamada93,Takahashi95, Takahashi98} and (Ca,Sr)Co$_{2}$P$_{2}$ \cite{Imai14} suggest a close connection between itinerant-electron metamagnetism and itinerant magnetic frustration.

\section{Acknowledgments}
This research was supported by the U.S. Department of Energy, Office of Basic Energy Sciences, Division of Materials Sciences and Engineering. Ames Laboratory is operated for the U.S. Department of Energy by Iowa State University under Contract No. DE-AC02-07CH11358.  A portion of this research used resources at the Spallation Neutron Source, a DOE Office of Science User Facility operated by the Oak Ridge National Laboratory.

\appendix
 
 \section{Analysis of INS data with the diffusive model}\label{INS_diff_anal}
\begin{figure*}[]
 	\centering
 	\includegraphics[scale=1.0]{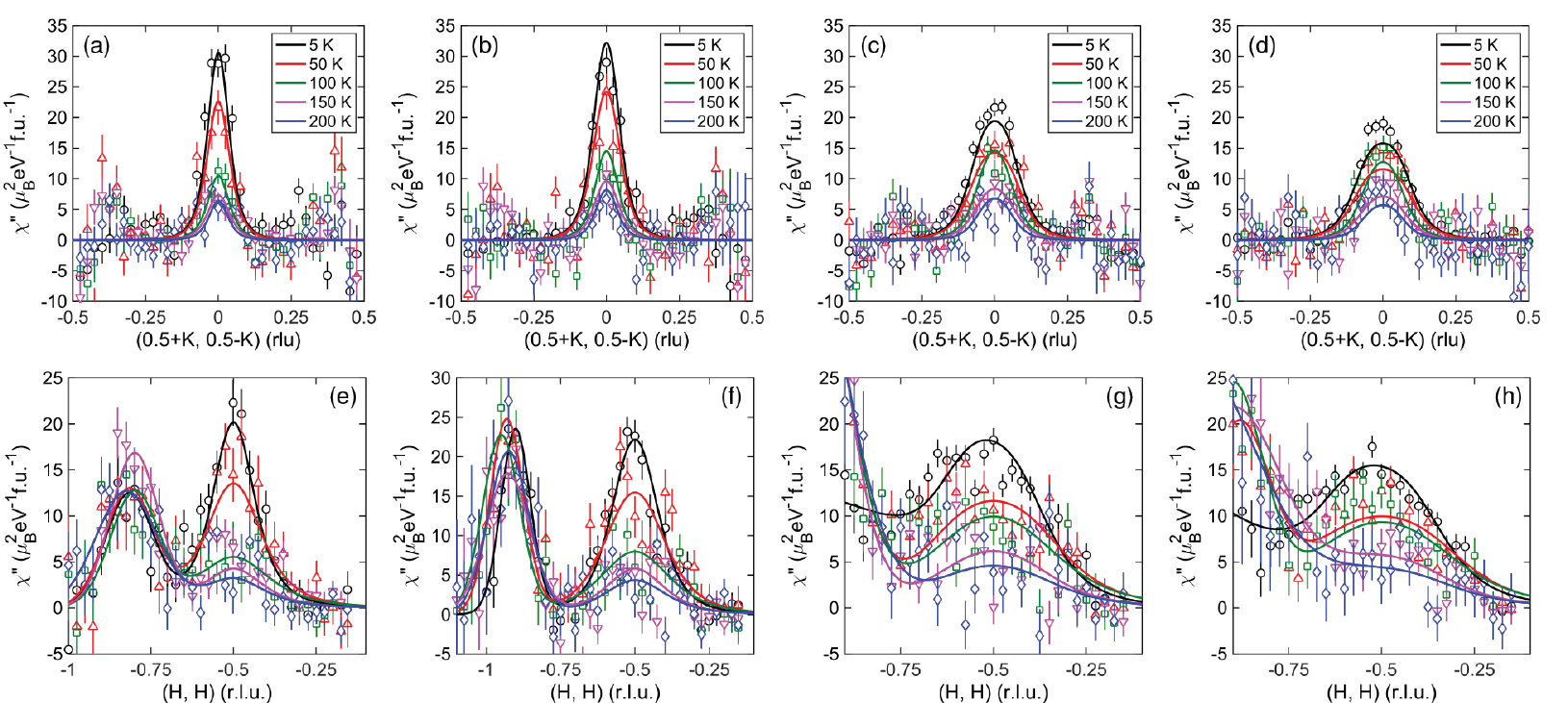}
 	\caption{(a-d) Transverse (TR) and (e-h) longitudinal (LO) cuts for $E = 5$ to $10$, $10$ to $15$, $30$ to $40$ and $40$ to $50$~meV. The TR cuts are averaged over $\pm0.1$~rlu in the LO direction, the LO cuts are averaged over $\pm0.1$~rlu in the TR direction.  Data are for an incident neutron energy of $E_{\text{i}}=75$~meV.}
 	\label{cuts}
 \end{figure*}

 Transverse (TR) and longitudinal (LO) cuts through $\mathbf{Q}_{\text{stripe}}$ for energy transfer ranges of $E = 5$ to $10$, $10$ to $15$, $30$ to $40$, and $40$ to $50$~meV, where the magnetic scattering largely avoids phonon scattering, are shown in Fig.~\ref{cuts}.  These cuts and the cuts in Fig.~\ref{fig_tdep}(d) were simultaneously fit to Eq.~\eqref{eq_diffusive} to determine the fitted parameters plotted in Fig.~\ref{fig_fits}.
   
\section{Estimation of $\bm{J_{\text{avg}}}$ from inelastic neutron scattering data}\label{INS_Jav}
Equation~\eqref{eq_vtr} is used to estimate $J_{\text{avg}}$ from the steep spin-wave velocity in the direction transverse to $\mathbf{Q}_{\text{stripe}}$ shown, for example, in Fig.~\ref{fig_tdep}(a).  Within linear spin-wave theory,
\begin{equation}
\begin{split}
v_{\text{TR}}&=\sqrt{2}aS\sqrt{4J_{2}^{2}-J_{1}^{2}}\\*&=2\sqrt{2}aSJ_{2}\sqrt{1-\eta^{2}},
\end{split}
\end{equation}
where $a$ is the lattice parameter of the $I4/mmm$ crystallographic unit cell and $\eta$ is defined in Eq.~\eqref{eq_eta_J1J2}.  Thus,
\begin{equation}
J_{2}=\frac{v_{\text{TR}}}{2\sqrt{2}aS\sqrt{1-\eta^{2}}}.
\end{equation}
Given that $J_{\text{avg}}^{2} = J_{1}^{2}+J_{2}^{2}$, we can write
\begin{equation}
J_{2}=\frac{J_{\text{avg}}}{\sqrt{1+4\eta^{2}}},
\end{equation}
and obtain
\begin{equation}
J_{\text{avg}}=\frac{v_{\text{TR}}}{2\sqrt{2}aS}\sqrt{\frac{1+4\eta^{2}}{1-\eta^{2}}}.
\end{equation}
Using this relation, $\eta=-0.63$ (Table~\ref{table_exp_calc}), and $v_{\text{TR}}=250$~meV\AA\ [Eq.~\eqref{eq_vtr}]; we find a lower bound for the magnetic energy scale of $J_{\text{avg}} \approx 75$~meV.
     
\section{Random-phase approximation to the $\bm{J_1}$-$\bm{J_2}$ Heisenberg model and scaling relations}\label{MF_scale}

The magnetic susceptibility $\chi(\mathbf{Q},E=0)$ in a random-phase approximation (RPA) at $\mathbf{Q}$ corresponding to the magnetic ordering propagation vector $\bm{\tau}$ for a local-moment system is \cite{White07}
\begin{equation}
\chi(\bm{\tau}, 0)=\frac{C}{T-T_{\text{N}}}, \label{eq_RPA_Curie}
\end{equation}
where $C$ is the Curie constant given by 
\begin{equation}
C=g^{2}\mu_{\text{B}}^{2}S(S+1)/3k_\text{B},
\end{equation}
$g$ is the spectroscopic splitting factor, $S$ is the spin of the magnetic ion, and
\begin{equation}
 T_{\text{N}}=S(S+1)J(\bm{\tau})/3k_\text{B}
 \end{equation}
is the N\'{e}el temperature.  Note that $T_{\text{N}}$ is distinct from the Weiss temperature $\theta$ for the uniform ($\mathbf{Q}=\bm{0}$) susceptibility: 
\begin{equation}
\theta=S(S+1)J(\mathbf{Q}=\bm{0})/3k_\text{B}.
\end{equation}
Substituting for $T_{\text{N}}$, Eq.~\eqref{eq_RPA_Curie} may be written as:
\begin{equation}
\chi(\mathbf{Q},0)=C\left[T-\frac{S(S+1)}{3k_\text{B}}J(\mathbf{Q})\right]^{-1}. \label{eq_MFchi_Q}
\end{equation}

For the $J_{1}$-$J_{2}$ model appropriate for the square-Co sublattice in the $I4/mmm$ unit cell of the ThCr$_{2}$Si$_{2}$ structure with lattice parameter $a$, the $\mathbf{Q}$-dependent exchange interaction is
\begin{equation}
\begin{split}
J(\mathbf{Q}) =-2J_{1}\{\cos{[\tfrac{a}{2}(Q_{x}+Q_{y})]}\\*+\cos{[\tfrac{a}{2}(Q_{x}-Q_{y})]\}}\\*-2J_{2}[\cos{(Q_{x}a)}+\cos{(Q_{y}a)}],
\end{split}
\end{equation}
where the subscripts $x$ and $y$ correspond to perpendicular directions connecting NN Co, and $J>0$ corresponds to AF interactions.  For this model, the uniform susceptibility is
\begin{equation}
\chi(\bm{0},0)=\frac{C}{T-\theta},
\end{equation}
with a Weiss temperature given by
\begin{equation}
\theta=-\frac{4(J_{1}+J_{2})S(S+1)}{3k_{\text{B}}}.
\end{equation}

To study the critical behavior near $\mathbf{Q}_{\text{stripe}}=\frac{2\pi}{a}(\frac{1}{2},\frac{1}{2})$, we expand around $\mathbf{Q}_{\text{stripe}}$:
\begin{gather}
\begin{split}
J(\mathbf{Q}_{\text{stripe}}+\mathbf{q})=-2J_{1}\{-\cos{[\tfrac{a}{2}(q_{x}+q_{y})]}\\+\cos{[\tfrac{a}{2}(q_{x}-q_{y})]} \}\\+2J_{2}[\cos{(q_{x}a)}+\cos{(q_{y}a)}],
\end{split}\nonumber\\ \ \
\begin{split}
\phantom{J(\mathbf{Q}_{\text{stripe}}+\mathbf{q})}\approx-2J_{1}[\tfrac{1}{2}(\tfrac{a}{2})^{2}(q_{x}+q_{y})^{2}\\-\tfrac{1}{2}(\tfrac{a}{2})^{2}(q_{x}-q_{y})^{2}]\\+2J_{2}[2-\tfrac{1}{2}(q_{x}a)^{2}-\tfrac{1}{2}(q_{y}a)^{2}],
\end{split}\nonumber
\intertext{which gives}
J(\mathbf{Q}_{\text{stripe}}+\mathbf{q})\approx 4J_{2}-J_{2}a^{2}q^{2}-J_{1}a^{2}q_{x}q_{y}.
\end{gather}
We then obtain the static susceptibility near $\mathbf{Q}_{\text{stripe}}$:
\begin{multline}
\chi(\mathbf{Q}_{\text{stripe}}+\mathbf{q},0)=\\*C\left[T-\frac{S(S+1)}{3k_{\text{B}}}(4J_{2}-J_{2}q^{2}a^{2}-J_{1}a^{2}q_{x}q_{y})\right]^{-1}.
\end{multline}
We identify
\begin{equation}
T_{\text{N}}=\frac{4S(S+1)J_{2}}{3k_\text{B}},
\end{equation}
and write
\begin{multline}
\chi(\mathbf{Q}_{\text{stripe}}+\mathbf{q},0)=\\*\chi(\mathbf{Q}_{\text{stripe}},0)\left[1+\frac{J_{2}a^{2}S(S+1)}{3k_{\text{B}}(T-T_{\text{N}})}(q^{2}+2\eta q_{x}q_{y})\right]^{-1},
\label{etaJ1J2}
\end{multline}
where $\eta=J_{1}/(2J_{2})$, as given by Eq.~\eqref{eq_eta_J1J2}.

To connect to the $E=0$ diffusive susceptibility, we realize that $\chi(\mathbf{Q}_{\text{stripe}},0)=\chi^{\prime}(\mathbf{Q}_{\text{stripe}},0)$ and define the temperature-dependent correlation length
\begin{equation}
\begin{split}
\xi_{T}^{2}&=\frac{J_{2}a^{2}S(S+1)}{3k_\text{B}(T-T_{\text{N}})}\\* &=\xi_{0}^{2}\frac{T_{\text{N}}}{T-T_{\text{N}}},
\end{split}
\end{equation}
where $\xi_{0}=a/2$.

The susceptibility can now be written in the $E=0$ diffusive form as
\begin{equation}
\chi^{\prime}(\mathbf{Q}_{\text{stripe}}+\mathbf{q},0)=\frac{\chi^{\prime}(\mathbf{Q}_{\text{stripe}},0)}{1+\xi_{T}^{2}(q^{2}+2\eta q_{x}q_{y})},
\end{equation}
and we define a scaling relation between the static susceptibility and the correlation length within the RPA: 
\begin{equation}
\frac{\chi^{\prime}(\mathbf{Q}_{\text{stripe}},0)}{(\xi_{T}/a)^{2}}=\frac{g^{2}\mu_{\text{B}}^{2}}{J_{2}}.
\end{equation}
We next write \begin{equation}
\chi^{\prime}(\mathbf{Q}_{\text{stripe}}+\mathbf{q},0)=\frac{\chi_{s}}{(\xi_{0}/\xi_{T})^{2}+\xi_{0}^{2}(q^{2}+2\eta q_{x}q_{y})},
\end{equation}
where the bare staggered susceptibility, $\chi_{s}$, is
\begin{align}
\chi_{s}&=\frac{\chi^{\prime}(\mathbf{Q}_{\text{stripe}},0)}{(\xi_{T}/\xi_{0})^{2}}\nonumber\\&=\frac{g^{2}\mu_{\text B}^{2}}{4J_{2}}.\label{C14}
\end{align}
With this definition, we now recast the Curie-Weiss susceptibility in terms of the bare staggered susceptibility as
\begin{equation}
\chi^{\prime}(\mathbf{Q}_{\text{stripe}},0)=\chi_{s}\frac{T_{\text{N}}}{T-T_{\text{N}}},
\end{equation}
which gives Eq.~\eqref{eq_E0_RPA_Curie}.

\bibliographystyle{apsrev4-1.bst}
\bibliography{SrCo2As2_INS_Tdep.bib}

\end{document}